\documentclass[12pt,preprint]{emulateapj}

\begin{document}
\title{Extremely Small Sizes for Faint $z\sim2$-8 Galaxies in the {\it
    Hubble} Frontier Fields: A Key Input For Establishing their Volume
  Density and UV Emissivity} \author{R.J. Bouwens\altaffilmark{1},
  G.D. Illingworth\altaffilmark{2}, P.A. Oesch\altaffilmark{3,4},
  H. Atek\altaffilmark{3}, D. Lam\altaffilmark{1},
  M. Stefanon\altaffilmark{1}} \altaffiltext{1}{Leiden Observatory,
  Leiden University, NL-2300 RA Leiden, Netherlands}
\altaffiltext{2}{UCO/Lick Observatory, University of California, Santa
  Cruz, CA 95064} \altaffiltext{3}{Department of Astronomy, Yale
  University, New Haven, CT 06520}
\altaffiltext{4}{Observatoire de Gen{\`e}ve, 1290 Versoix, Switzerland}
\begin{abstract}
We provide the first observational constraints on the sizes of the
faintest galaxies lensed by the {\it Hubble} Frontier Fields (HFF)
clusters.  Ionizing radiation from faint galaxies likely drives cosmic
reionization, and the HFF initiative provides a key opportunity to
find such galaxies. Yet, we cannot assess their ionizing emissivity
without a robust measurement of their sizes, since this is key to
quantifying both their prevalence and the faint-end slope to the $UV$
luminosity function.  Here we provide the first size constraints with
two new techniques. The first utilizes the fact that the detectability
of highly-magnified galaxies as a function of shear is very dependent
on a galaxy's size.  Only the most compact galaxies remain detectable
in high-shear regions (vs. a larger detectable size range for low
shear), a phenomenon we quantify using simulations.  Remarkably,
however, no correlation is found between the surface density of faint
galaxies and the predicted shear, using 87 high-magnification
($\mu=10$-100) $z\sim2$-8 galaxies seen behind the first four HFF
clusters.  This can only be the case if faint ($\sim-$15 mag) galaxies
have significantly smaller sizes than more luminous galaxies, i.e.,
$\lesssim$30 mas or 160-240 pc. As a second size probe, we rotate and
stack 26 faint high-magnification sources along the major shear axis.
Less elongation is found than even for objects with an intrinsic
half-light radius of 10 mas. Together these results indicate that
extremely faint $z\sim2$-8 galaxies have near point-source profiles
(half-light radii $<$30 mas and perhaps 5-10 mas). These results
suggest smaller completeness corrections and hence shallower faint-end
slopes for the $z\sim2$-8 LFs than derived in some recent studies (by
$\Delta\alpha\gtrsim0.1$-0.3).
\end{abstract}

\section{Introduction}

Over the last few years, there has been increasing interest in the
study of faint galaxies in the high-redshift universe, both for
guiding current thinking about the reionization of the universe
(Kuhlen \& Faucher-Gigu{\'e}re 2012; Robertson et al.\ 2013;
Choudhury et al.\ 2015) and also for the interpretation of dwarf
galaxies much more locally (e.g., Graus et al.\ 2016).  The importance
of faint galaxies to cosmic reionization follows from the strong
observational evidence that the faint-end slope of the $UV$ LF is as
steep as $\sim-2$ at $z>5$ (e.g., Yan \& Windhorst 2004; Bouwens et
al.\ 2007, 2011, 2015; Oesch et al.\ 2012; Bradley et al.\ 2012; Calvi
et al.\ 2013; Schenker et al.\ 2013; McLure et al.\ 2013; Schmidt et
al.\ 2014; Finkelstein et al.\ 2015; Atek et al.\ 2014, 2015a, 2015b;
Castellano et al.\ 2016b), implying that the vast majority of
high-energy $UV$ photons originate from the extremely faint galaxies.

Substantial progress has been made in pushing fainter in searches for
faint galaxies in the early universe.  Traditionally, our deepest
probes have been provided by the long exposures obtained over the {\it
  Hubble} Ultra Deep Field (HUDF: Beckwith et al.\ 2006).  Searches
over this field first probed to $\sim-17.7$ mag (Bouwens et al.\ 2011;
Oesch et al.\ 2012; Bradley et al.\ 2012) at $z\sim7$-8 and later to
$\sim-17$ (Schenker et al.\ 2013; McLure et al.\ 2013; Bouwens et
al.\ 2015), in the HUDF/XDF/HUDF12 (Illingworth et al.\ 2013;
Koekemoer et al.\ 2013).  Bouwens et al.\ (2015) probe to $\sim-15.8$
mag at $z\sim4$, and Parsa et al.\ (2016) take advantage of the
smaller luminosity distances at $z\sim2$-3 to reach $\sim-$14 mag.

Over the last few years, however, the effort to identify extremely
faint galaxies has been given a major boost, due to the new 840-orbit
{\it Hubble} Frontier Fields (HFF) program (Coe et al.\ 2015; Lotz et
al.\ 2017).  This program probes faint galaxies by combining the power
of flux amplification by gravitational lensing from massive galaxy
clusters with long exposures by the {\it Hubble Space Telescope} and
other telescopes.  Many researchers have exploited new observations
from this program to study faint galaxies.  Atek et al.\ (2014, 2015)
were the first to make use of observations from this program and some
of their first results probed as faint as $-15$ mag.  New searches by
Kawamata et al.\ (2016), Castellano et al.\ (2016a,b), and Livermore
et al.\ (2017) now report the identification of $z\sim5$-6 galaxies as
faint as $\sim-13$ mag.

\begin{figure*}
\epsscale{1.16}
\plotone{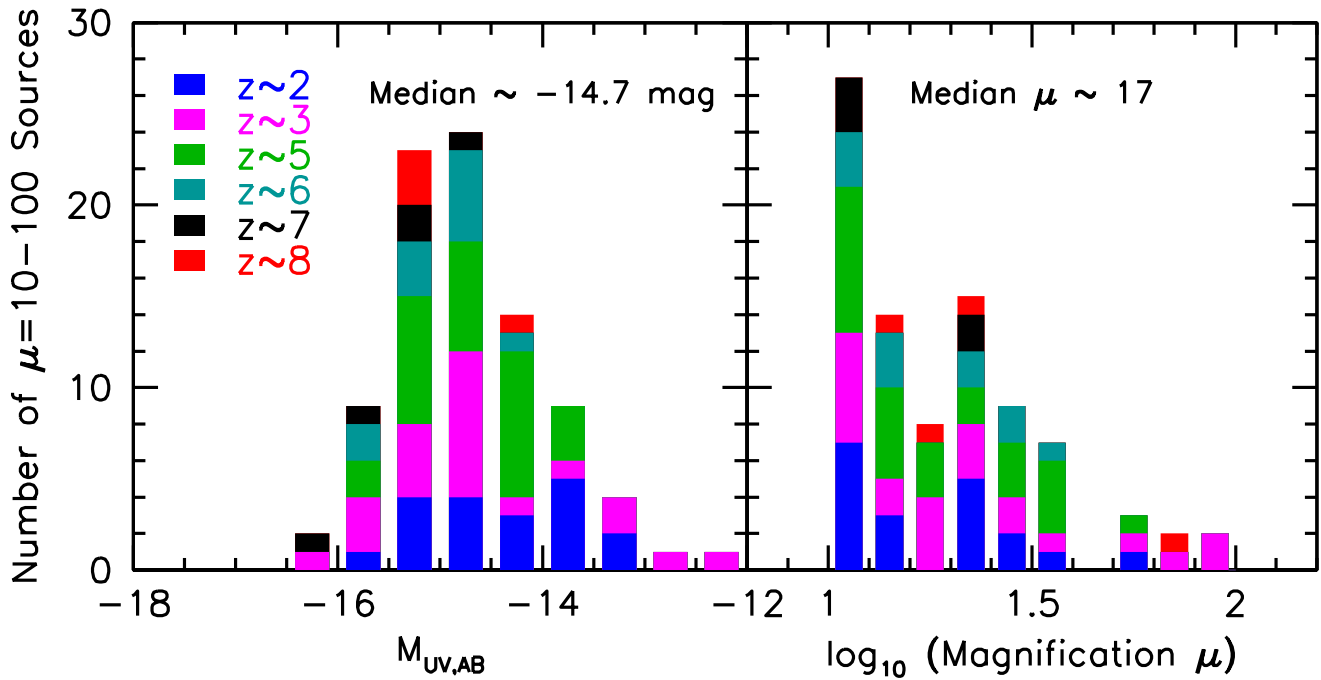}
\caption{Number of high-magnification ($\mu=10$-100) $z\sim2$,
  $z\sim3$, $z\sim5$, $z\sim6$, $z\sim7$, and $z\sim8$ sources
  identified over the 4 HFF clusters considered here vs. absolute
  magnitude $M_{UV}$ (\textit{left}) and estimated magnification $\mu$
  (\textit{right}).  The estimated magnification we utilize is the
  median of the four public parametric models for the HFF clusters.
  We only include sources where the median magnification estimate from
  the parametric models is not more than double the geometric mean of
  the lowest 2 magnification estimates from the parametric models.
  Also the sample is restricted to sources with apparent magnitudes
  $>$28 mag at $z=5$-8 and $>$26.5 mag at $z=2$-3 to focus on the
  properties of lowest luminosity sources.\label{fig:hist}}
\end{figure*}

The HFF observations therefore have great potential for mapping out
the faint-end of the luminosity function (LF).  Nevertheless, there
are a number of issues that make such LF derivations more challenging
than LF determinations using traditional deep field observations like
the HUDF.  Among these issues are (1) the gravitational lens model
utilized to determine the luminosity of faint sources, (2) the size
distribution of faint sources needed to estimate the selection
volumes, and (3) possible contamination from foreground sources in the
cluster.  Each issue has a host of uncertainties associated with it,
and if not treated correctly, each can result in sizeable systematic
errors.

\begin{deluxetable}{ccccccc}
\tablewidth{0cm}
\tablecolumns{7}
\tabletypesize{\footnotesize}
\tablecaption{Samples of High-Magnification $\mu=10$-100 $z=2$-8 Galaxies Found over the First Four HFF clusters\tablenotemark{a}\label{tab:samp}}
\tablehead{\colhead{Cluster} & \colhead{$z$$\sim$2} & \colhead{$z$$\sim$3} & \colhead{$z$$\sim$5} & \colhead{$z$$\sim$6} & \colhead{$z$$\sim$7} & \colhead{$z$$\sim$8}}
\startdata
Abell 2744 & 2 & 10 & 8 & 1 & 1 & 2 \\
MACS0416 & -- & -- & 6 & 7 & 1 & 0 \\
MACS0717 & 15 & 10 & 10 & 2 & 3 & 2 \\
MACS1149 & 2 & 2 & 2 & 1 & 0 & 0 \\
Total & 19 & 22 & 26 & 11 & 5 & 4 
\enddata
\tablenotetext{a}{Samples to be presented in Table~\ref{tab:catalog}.  See \S2}
\end{deluxetable}

Here we focus on the issue of source size for extremely faint galaxies
seen behind the HFF clusters.  Source size is known to have a very
large impact on the estimated selection efficiencies near the
detection limits and hence inferred volume densities.  Grazian et
al.\ (2011) highlighted source size as having a substantial impact on
the faint-end slope $\alpha$ inferred for the $UV$ LF, arguing that
the faint-end slope $\alpha$ derived can be significantly dependent on
assumptions made regarding source size.  This issue is important even
for nominally small sources (i.e., $\lesssim$0.5 kpc) in cases where
lensing magnification becomes significant.  This is due to the
substantial stretching sources experience as a result of gravitational
lensing.  This can make magnified sources difficult to detect, even,
if from a consideration of their flux, detection should be
straightforward.  This issue is particularly problematic if lensing
acts to stretch their light predominantly along a single axis (see
Oesch et al.\ 2015).

The purpose of this manuscript is to demonstrate the application of
several new techniques to constrain the size distribution of faint
galaxies identified behind lensing clusters.  The first technique keys
on the expectation that the search efficiency behind lensing clusters
should be highest in regions where sources are magnified with minimal
shear and lowest in regions where the lensing shear is high.  Given
that we would expect there to be the largest differences between these
regimes in cases where galaxy sizes are large and essentially no
difference in cases of a point-source profile, this strategy provides
us with a valuable way of estimating source size for faint galaxies.
With our second technique, we obtain our size constraints by looking
at highly magnified sources stretched by $>$10$\times$ along a single
axis and then comparing their profiles with expectations based on
current lensing models.

The plan for this paper is as follows.  We begin the paper by
introducing the data sets and samples we will be using to look at the
issue of source size (\S2).  We then move on to illustrate the impact
that the assumed source size can have on the inferred UV LF (\S3).  In
\S4, we use simulations to investigate how the completeness of
high-magnification, faint galaxies should depend on lensing shear for
a variety of different assumptions about source size and then look for
similar dependencies in the observations.  In \S5, we obtain a
constraint on source size by looking at a selection of faint sources
expected to be stretched by a factor of $>$10 along a single axis and
then comparing their spatial profiles with those expected from the
lensing models.  In \S5, we also direct size measurements from a
sample of faint $z\sim6$ sources behind Abell 2744 and MACS0416.
Finally, in \S6 and \S7, we discuss and summarize the results.
Throughout the paper, we assume a standard ``concordance'' cosmology
with $H_0=70$ km s$^{-1}$ Mpc$^{-1}$, $\Omega_{\rm m}=0.3$ and
$\Omega_{\Lambda}=0.7$, which is in good agreement with recent
cosmological constraints (Planck Collaboration et al.\ 2015).
Magnitudes are in the AB system (Oke \& Gunn 1983).

\section{Data Sets and $z=2$-8 Samples}

We base the present study on the v1.0 reductions of the {\it HST}
observations over the first four HFF clusters Abell 2744, MACS0416,
MACS0717, and MACS1149 (Koekemoer et al.\ 2017, in prep).  These data
include at least 18, 10, 42, 34, 12, 10, and 24 orbits of F435W,
F606W, F814W, F105W, F125W, F140W, and F160W observations,
respectively, typically probing to $\sim$28.8-29.1 mag at $5\sigma$
for point sources (Lotz et al.\ 2017).  The FWHM of the PSF in the
F105W, F125W, F140W, and F160W WFC3/IR observations is typically
$\sim0.16$-0.17''.

We also make use of our own reduction of the {\it HST} WFC3/UVIS F275W
and F336W observations available over Abell 2744, MACS0717, and
MACS1149.  These observations include 8 orbits of data in the F275W
and F336W bands and reach to a depth of $\sim$27.4-28.2 mag (Alavi et
al.\ 2016).  These observations help us to construct samples of very
faint galaxies at $z\sim2$ and $z\sim3$ which we will also use to
study galaxy sizes.

We consider a conservative yet comprehensive selection of $z\sim2$-8
galaxies identified over the first four HFF clusters.  Foreground
light from the cluster and the 40 brightest galaxies have been removed
from the real data before combination with the simulated data.  Our
procedure for removing the foreground light relies both on {\sc
  galfit} (Peng et al.\ 2002) and the median filtering approach from
SExtractor (Bertin \& Arnouts 1996) applied at two grid scales; our
procedure shares many similarities with the approach taken by Merlin
et al.\ (2016: see Bouwens et al.\ 2017, in prep).  \textsc{galfit} is
a well-known two dimensional profile fitting code that produces robust
size measurements from imaging observations, given an input PSF.

\begin{figure}
\epsscale{1.16}
\plotone{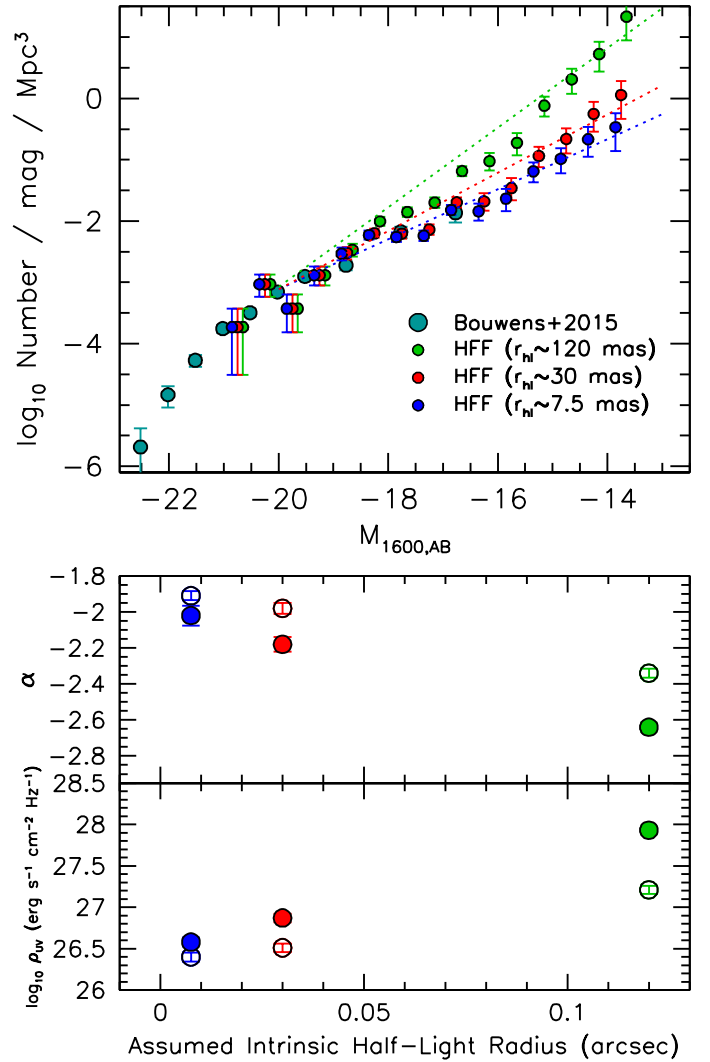}
\caption{(\textit{upper}) Three different determinations of the
  $z\sim6$ LF (circles with $1\sigma$ error bars) adopting different
  assumptions about the size of faint $z\sim6$ galaxies.  The green,
  red, and blue circles assume lognormal size distributions with a
  $r_{hl}$ $\sim$ 120 mas, 30 mas, and 7.5 mas (unlensed),
  respectively, for faint galaxies, with a $1\sigma$ scatter of 0.3
  dex.  The points have been offset horizontally for clarity.
  (\textit{lower two panels}) The lower two panels show the faint-end
  slopes and $UV$ luminosity densities (integrated to $-$13 mag) that
  one infers for the $UV$ LF at $z\sim6$ derived using the different
  size assumptions.  Faint-end slope results are shown (open and solid
  circles) fitting to the brighter ($<-15$) and fainter ($>-15$)
  lensed LF results, respectively, with the implied $UV$ luminosities
  shown for the faint-end slope results shown with open and solid
  circles, respectively.  Clearly, assumptions about source size can
  have a huge impact on the volume density of faint galaxies inferred
  from the HFF program.  The effective faint-end slopes $\alpha$ of
  the green and blue LFs differ by $\Delta\alpha\sim0.75$ and the $UV$
  luminosity densities inferred differ by a factor of
  40.\label{fig:lf6size}}
\end{figure}

Our $z\sim2$-8 galaxy candidates were selected by a combination of the
Lyman Break and photometric-redshift selection criteria.  We will
describe those selection criteria and the $z\sim2$-8 samples we
construct in detail in Bouwens et al.\ (2017, in prep), but the
criteria we utilize are almost identical to those utilized in Bouwens
et al.\ (2015) for our $z\sim5$-8 samples and involve
photometric-redshift selection criteria at $z\sim2$-3.  Selected
sources are required to be detected at 6.5$\sigma$ adding in
quadrature the S/N of each source in the $Y_{105}$, $J_{125}$,
$JH_{140}$, and $H_{160}$ bands to guarantee a clean selection of
sources.  No selection of $z\sim4$ galaxies is considered due to
potentially significant contamination of such samples by sources at
the redshift of the cluster with 4000\AA/Balmer breaks falling between
the $B_{435}$ and $V_{606}$ bands.

\begin{figure*}
\epsscale{1.15}
\plotone{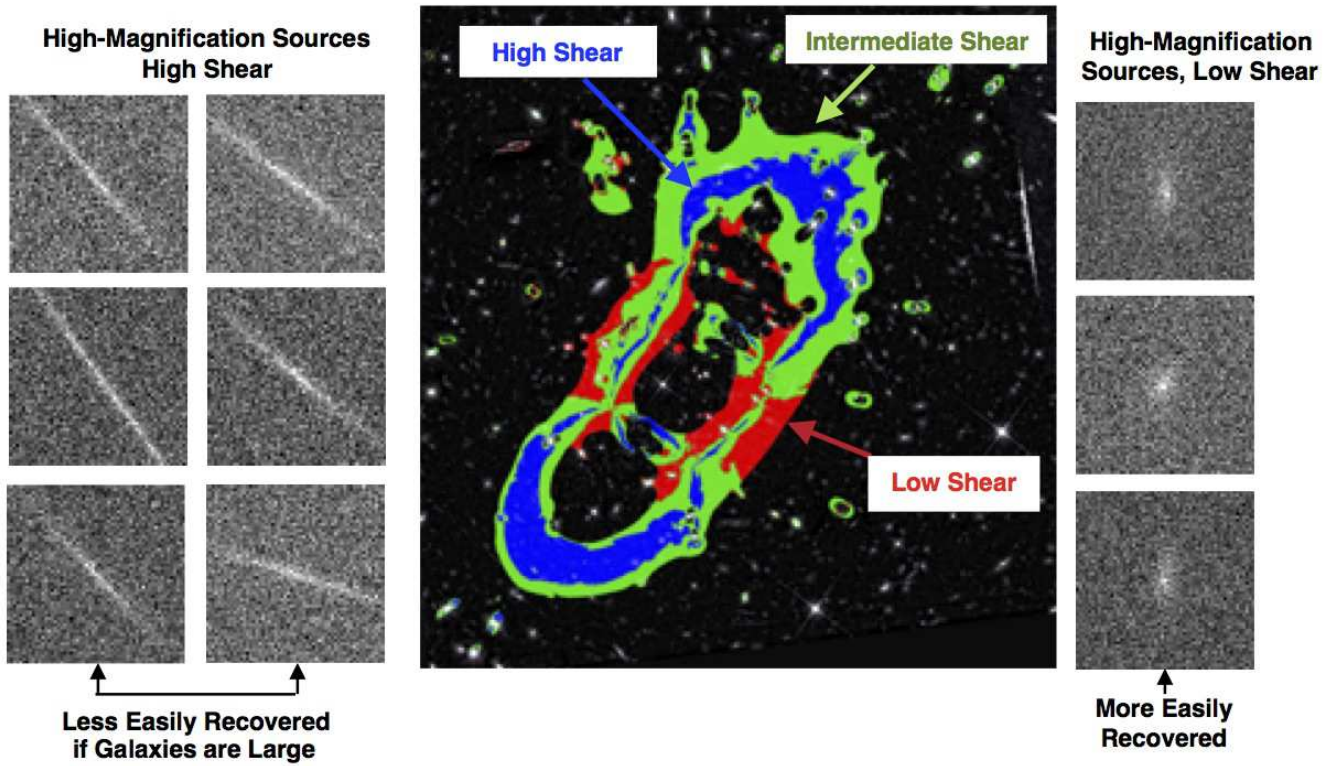}
\caption{(\textit{center}) Image of the HFF cluster Abell 2744.
  High-magnification ($\mu=10$-100) regions are explicitly indicated,
  with red, green, and blue shading for sources where the predicted
  shear factors $S$ (see Eq. 1) are expected to be low ($<2.5$),
  intermediate (2.5-10), and high ($>10$), respectively.
  (\textit{left} and \textit{right}) A few zoom-in images of model
  sources are shown, to provide readers with an illustration of the
  expected morphologies of sources located in various regions around
  the cluster.  The maximum surface brightness of sources is kept
  similar to allow for consistency visibility of the sources, and no
  convolution of the zoom-in images with a WFC3/IR PSF is considered
  (to ensure the impact of the lensing distortion on the morphologies
  is clear).}
\end{figure*}

In total, 559 $z\sim2$, 562 $z\sim3$, 309 $z\sim5$, 160 $z\sim6$, 92
$z\sim7$, and 50 $z\sim8$ galaxies were selected.  $\sim$5\% of these
candidates are estimated to have magnification factors $>$10, using
the median of the four public magnification models to the HFF program
based on parametric NFW mass profiles, i.e., CATS (Jullo \& Kneib
2009; Richard et al.\ 2014; Jauzac et al.\ 2015a,b), Sharon (Johnson
et al.\ 2014), GLAFIC (Oguri 2010; Ishigaki et al.\ 2015; Kawamata et
al.\ 2016), and Zitrin-NFW (Zitrin et al.\ 2013, 2015).  These models
performed the best in the HFF comparison project (Meneghetti et
al.\ 2016).  The number of high-magnification sources per cluster and
in different bins in magnitude and magnification factor are provided
Table~\ref{tab:samp} and Figure~\ref{fig:hist}.

\section{Importance of Source Size For Constraints on the Faint End of 
the $UV$ LFs at $z\geq 2$}

It is useful first to provide some perspective on the importance of
the assumed size distribution for determinations of the rest-frame
$UV$ LF.  We illustrate the impact at just one redshift $z=6$, due to
the importance of this redshift for current thinking about cosmic
reionization and the fact that HFF observations are the most sensitive
in those passbands which straddle the $z\sim6$ Lyman break.

To demonstrate the effect of source size, we consider three different
size assumptions for faint $z\sim6$ galaxies behind the HFF clusters.
These are chosen to differ in size by a factor 4 at each step,
starting with a typical size for brighter galaxies (around $L^*$ in
brightness) of (1) 0.12$''$ (120 mas), and then taking (2) 30 mas and
(3) 7.5 mas.  More specifically the actual size distributions assumed
are as follows: (1) log-normal with median half-light radii of
$\sim$120 mas (unlensed) and 1$\sigma$ scatter of 0.3 dex, (2)
log-normal with median half-light radii of $\sim$30 mas (unlensed) and
1$\sigma$ scatter of 0.3 dex, and (3) a delta function with a peak at
a half-light radius of 7.5 mas.

The $z\sim6$ LFs we derive for this exercise take advantage of
selection volumes we have estimated based on sophisticated image
construction and recovery simulations.  These simulations involve
first creating a mock catalog of sources, simulating the appearance of
these galaxies in the source plane, mapping these sources to the image
plane using one current state-of-the-art lensing model (which we take
to be CATS: Jauzac et al.\ 2015), convolving with the relevant point
spread function, adding the mock image plane observations to the real
data, and then running the present detection and source selection
algorithms on the mock data in the same way as it was run on the real
observations.

In simulating the appearance of sources at all wavelengths, we assume
that the mean $UV$-continuum slope $\beta$ of galaxies matched the
constraints available in Bouwens et al.\ (2014), where the $\beta$'s
are redder $\sim-1.5$ for the more luminous galaxies and bluer
$\sim-2.3$ for the fainter sources.  The Bouwens et al.\ (2014)
constraints are broadly representative of those found in numerous
studies (Wilkins et al.\ 2011; Bouwens et al.\ 2012; Finkelstein et
al.\ 2012; Dunlop et al.\ 2013; Alavi et al.\ 2014; Rogers et
al.\ 2014; Duncan et al.\ 2014).  Sources are assumed to all have
exponential profiles, which is a rough match to the average profile of
many $z\sim4$ galaxies (Hathi et al.\ 2008; Shibuya et al.\ 2015).

Then, combining these selection efficiencies with a large sample of
$z\sim6$ galaxies presented in Bouwens et al.\ (2017: see
Table~\ref{tab:catalog}), we derive different estimates of the $UV$ LF
at $z\sim6$.  Figure~\ref{fig:lf6size} presents these estimates of the
$UV$ LFs, as well as the faint-end slopes $\alpha$.  We derive
separate LF fit results alternatively using the brighter ($<-15$ mag)
and fainter ($>-15$ mag) individual points in the LF.  Estimates of
the $UV$ luminosity density brightward of $-13$ mag for the derived LF
are shown for the different fit results.  For simplicity, the
normalization $\phi^*$ and $M^*$ are kept fixed to the values derived
in multi-field LF probe by Bouwens et al.\ (2015), i.e., $\phi^* =
0.50_{-0.16}^{+0.22} \times 10^{-3}$ Mpc$^{-3}$ and $M^* =
-20.94\pm0.20$.  Only sources brighter than 29 mag were included in
our LF derivations.  Our procedure for estimating the $UV$ LF is
described in more detail in Bouwens et al.\ (2017), but remains
similar to the procedures used in our previous extensive analyses
(e.g. Bouwens et al.\ 2015).

It is clear from Figure~\ref{fig:lf6size} that the assumed source
sizes can have a huge impact on the LFs, faint-end slopes $\alpha$,
and inferred $UV$ luminosity densities we derive, even when the
differences are as small as $\sim$7.5 mas vs. $\sim$30 mas.  This
motivates our attempts to accurately measure the size distribution of
extremely faint galaxies.

\section{Galaxy Sizes from Dependencies on the Lensing Shear}

\subsection{Formalism and Description of Technique}

We begin this section by introducing the lensing terminology we will
utilize to constrain the size distribution of highly-magnified, very
faint galaxies behind lensing clusters.

It is traditional in looking at the impact of gravitational lensing
from various mass distributions on light in the unlensed ``source''
plane to write the transformation to the lensed ``image'' plane in
terms in a linearized form using the Jacobian $\frac{\partial
  \beta_i}{\partial \theta_j}$ where $\beta$ and $\theta$ expresses
the unlensed angular and observed angular positions, respectively.

\begin{figure}
\epsscale{1.15}
\plotone{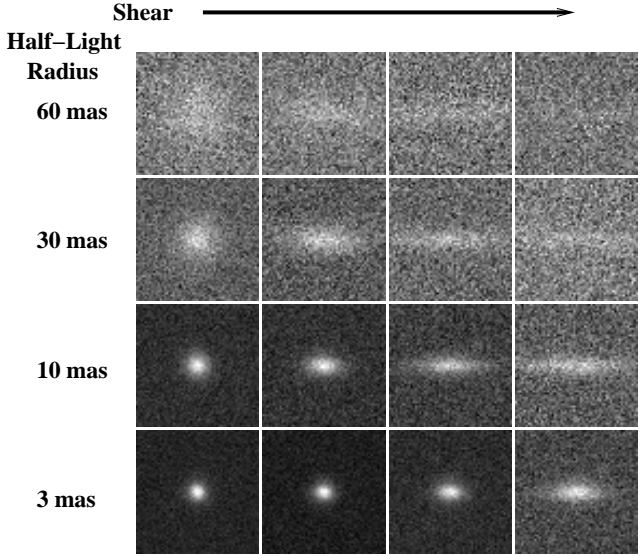}
\caption{Sources of fixed apparent magnitude and magnification factor
  ($\mu=20$) but with varying intrinsic half-light radii (60 mas, 30
  mas, 10 mas, 3 mas) and differing degrees of shear (with shear
  factors of 1, 5, 25, 125 from left to right).  These simulated
  images include the impact of the {\it HST} WFC3/IR PSF to make them
  fully realistic.  Sources subject to higher shear are much less
  easily selected than lower-shear sources, but the dependence of
  completeness on shear is a sensitive function of source size.  For
  sources with sizes $<$10 mas, the dependence on shear is less, such
  that we would expect the recovered surface densities of sources in
  low and high shear regions to be more similar.\label{fig:shearill}}
\end{figure}

\begin{figure*}
\epsscale{1.15}
\plotone{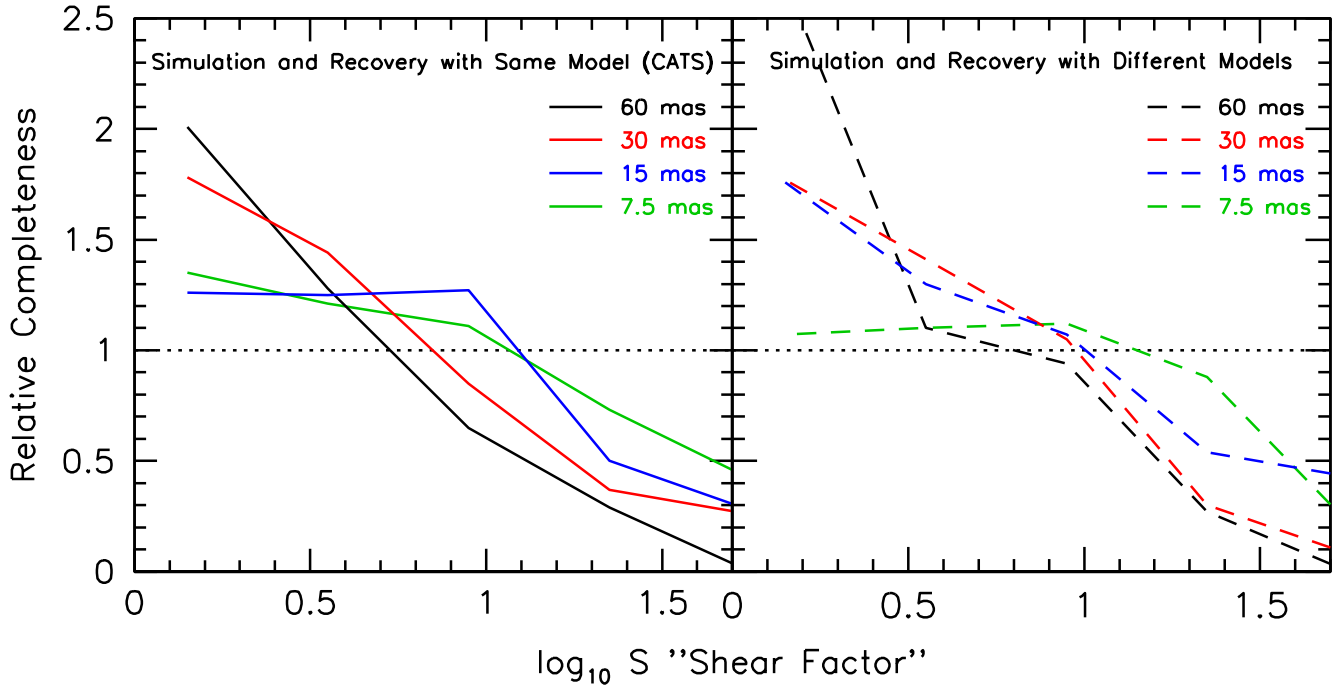}
% /Dropbox/faintend/z6_show/bin3.sm
\caption{Relative completeness of high-magnification sources expected
  as a function of the shear factor $S$ (see Eq. 1) assuming fixed
  half-light radii of 60 mas (\textit{black line}), 30 mas
  (\textit{red line}), 15 mas (\textit{blue line}), and 7.5 mas
  (\textit{green line}).  The derived completeness is computed only
  including the faintest $z\sim6$ galaxies in our fields, i.e., $>$28
  mag, and only for those sources with magnification factors $>$10;
  the completeness is functionally equivalent to the surface density
  of galaxies predicted to lie in various shear regimes.  The values
  plotted here for the completeness are normalized such that the
  average completeness (assuming an intrinsic half-light radius) is
  equal to 1 (so it is not possible to use this figure to compare the
  estimated completeness for two different assumptions about the
  size).  The left-hand panel shows the dependence as a function of
  the same shear field used in the simulations (\textit{solid lines}),
  while the right-hand panel also shows this dependence as a function
  of the median shear factor computed from seven high-resolution
  lensing maps available over the first four HFF clusters
  (\textit{dashed lines}).  Results are presented as a function of
  some median magnification map (to be distinct from the CATS lensing
  model used in the simulations) to illustrate how the dependencies on
  shear would change, if the evaluation was performed using different
  maps than were actually used in the simulations.  Source
  completeness is expected to be higher in regions where the shear is
  low (i.e., similar source elongation in both spatial dimensions)
  than in regions where the shear is high.  Dependencies on shear are
  weaker in cases where sources are intrinsically
  small.\label{fig:predcompshear}}
\end{figure*}

This transformation is frequently written in terms of the following
$2\times2$ matrix:
\begin{displaymath}
\left ( 
\begin{array}{cc}
1 - \kappa - \gamma_1 & \gamma_2 \\
\gamma_2 & 1 - \kappa + \gamma_1
\end{array}
\right)
\end{displaymath}
where $\kappa$ is the convergence while $\gamma_1$ and $\gamma_2$ are
the components of the shear.  One can choose the $x$ and $y$ axes such
that distortion transformation is diagonal:
\begin{displaymath}
\left ( 
\begin{array}{cc}
1 - \kappa - \gamma & 0 \\
0 & 1 - \kappa + \gamma
\end{array}
\right)
\end{displaymath}
$\kappa$ is equal to the surface density of matter divided by the
critical surface density $\Sigma_{cr}$ which is equal to $\frac{c^2
  D_s}{4\pi G D_{ls} D_{l}}$ where $c$ is the speed of light, $G$ is
Newton's constant, $D_s$, $D_{l}$, $D_{ls}$ are angular diameter
distances from the observer to the source, from the observer to the
lens, and from the lens to the source, respectively.  Meanwhile,
$\gamma$ is the shear.

A circular source of size $R$ in the source plane would have a
projected size of
\begin{displaymath}
\frac{R}{1-\kappa - \gamma}
\end{displaymath}
along its major axis and
\begin{displaymath}
\frac{R}{1-\kappa + \gamma}
\end{displaymath}
along its minor axis.  The resultant axis ratio of a circular source
would be as follows:
\begin{displaymath}
\frac{1-\kappa - \gamma}{1-\kappa+\gamma}
\end{displaymath}
We define a new quantity $S$, which we call the ``shear factor,'' and
take $S$ to equal the above expression in cases where it is greater or
equal to one and where it is reciprocal of the above expresson in
cases where it is less than one, i.e.,
\begin{equation}
S = \left\{
\begin{array}{lr}
\frac{1-\kappa - \gamma}{1-\kappa+\gamma}, &
\text{for } \frac{1-\kappa - \gamma}{1-\kappa+\gamma} \geq 1\\
\frac{1-\kappa + \gamma}{1-\kappa-\gamma}, &
\text{for } \frac{1-\kappa - \gamma}{1-\kappa+\gamma} < 1
\end{array}\right.
\end{equation}
For values of 1, sources would retain a circular shape, whereas for
values of $\sim$10, the axial ratio of lensed sources would be 10
(before accounting for the impact of the PSF).  The quantity $S$
expresses the shearing or spatial distortion of sources in the lensing
field.  Meanwhile, source magnification $\mu$ is simply equal to the
product of source stretch along the major and minor axes, i.e.,
\begin{displaymath}
\mu = \frac{1}{(1-\kappa)^2 - \gamma^2}
\end{displaymath}

Given current detection algorithms, we would expect a higher
completeness in regions of low shear compared to high shear regions
for a given apparent magnitude and magnification factor $\mu$ of
sources, particularly if the intrinsic sizes of high-redshift sources
is modest, i.e., $\sim$100 mas.  Sources elongated by similar factors
along the two spatial directions are easier to detect than sources
elongated predominantly along just one of the two spatial dimensions.
This is illustrated in Figure~\ref{fig:shearill} using sources with
various intrinsic sizes and subject to varying amounts of shear.  A
first discussion of the impact of this effect for finding faint
sources was provided by Oesch et al.\ (2015).

We would expect the strength of the dependence of surface density on
shear to vary in proportion to source size.  In fact, if we model
faint galaxies as point sources, the surface density of galaxies we
recover on the sky is entirely independent of the predicted shear and
is only a function of the magnification factor.  An illustration of
the reduced impact shear would have for smaller sources is evident in
Figure~\ref{fig:shearill} for the 3 mas case (which, even though
small, still clearly shows the reduction in detectability from shear).
This illustration motivates the systematic measurement of this
dependence from the data as a means of constraining the intrinsic
sizes of very faint high-redshift galaxies.

\subsection{Recovered Surface Density vs. Shear: Simulations}

Having described the basic principles that will be used in this
section and having illustrated the basic effect, we now use
simulations to quantify the expected dependence of completeness on the
predicted shear for sources of various sizes.  We focus on the
selection of $z\sim6$ galaxies in the magnitude interval $>28$ and
then discuss the extent that we might expect this selection of faint
$z\sim6$ galaxies to be representative of the selections at other
redshifts.

We accomplish this by running extensive source recovery simulations on
all 4 HFF clusters that we utilized to perform this basic test.
Briefly, we (i) populate the source plane with galaxies at some fixed
intrinsic magnitude, (ii) apply the deflection map from one recent
state-of-the-art lensing model (which we take to be the CATS models:
Jauzac et al.\ 2015), (iii) add the sources to the HFF data (after the
foreground cluster and brightest 50 cluster galaxy light has been
removed: see Bouwens et al.\ 2017, in prep), and (iv) then attempt to
identify $z\sim6$ galaxies using exactly the same procedure as was
used to originally select our high-redshift samples.  We repeat this
simulation hundreds of times systematically including as inputs a
different apparent magnitude for galaxies at random positions in the
source plane.

We present the results for Figure~\ref{fig:predcompshear}
alternatively assuming a fixed half-light radius of 60 mas, 30 mas, 15
mas, and 7.5 mas for distant $z\sim6$ galaxies (each of these radii
differing at the power of 2 level).  An intrinsic axial ratio of 1 is
adopted for sources in the simulations (i.e., all sources have an
intrinsically circular two dimensional profile).\footnote{This
  represents the typical case for sources, as the inclusion of
  non-circular sources in the simuations would either increase or
  decrease the completeness for an individiual source depending on
  whether the major axis is perpendicular or parallel, respectively,
  with the major shear axis.}  We only include sources
where the actual magnification is $>$10 and where the uncertainties on
the magnification is less than 0.3 dex (as determined by comparing the
1st quartile value with the median).  The shear factors we utilize are
derived from the CATS model.

As expected, we can see that our simulations find that sources
inserted into regions with low shear factors show a significantly
higher completeness than sources inserted into regions where the shear
is higher.  For our models where the source sizes are smaller, the
dependence of the completeness on the shear factor is less sharp.
Nevertheless, we do still observe a modest dependence, even for
sources with intrinsic half-light radii of 15 mas and 7.5 mas.

\begin{figure}
\epsscale{1.15}
\plotone{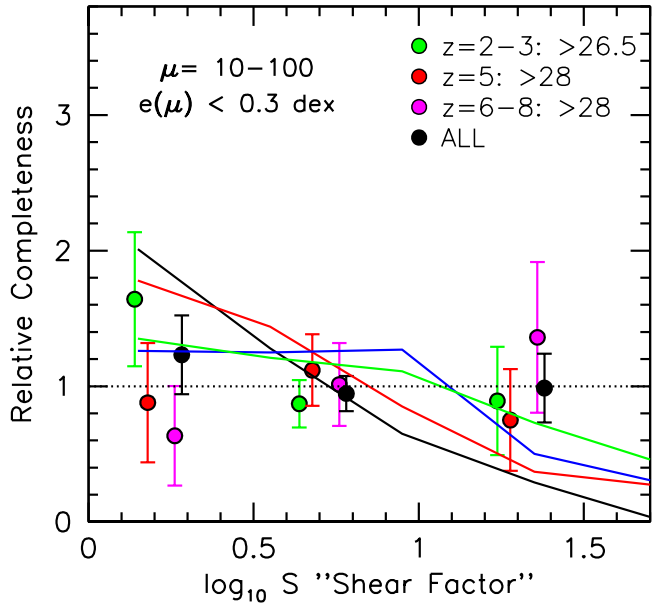}
% /Dropbox/lensed_samples/gencat/testlf/bin2.sm
\caption{Estimated relative completeness of high-magnification
  $\mu=10$-100 galaxies vs. shear factor $S$ (see Eq. 1).  Only
  galaxies faintward of $>$26.5 mag at $z\sim2$-3 and $>$28 at
  $z\sim5$-8 are included in the calculated surface densities.
  Results are shown for our $z\sim2$-3, $z\sim5$, and $z\sim6$-8
  samples individually (\textit{green, red, and magenta circles,
    respectively}) and for total sample of $z\sim2$-3 + $z\sim5$-8
  galaxies (\textit{black circles}).  In estimating the relative
  completeness vs. shear factor, we have assumed asymptotic faint-end
  slopes of $-1.45$, $-1.57$, $-1.81$, $-1.93$, $-2.05$, and $-2.17$
  for our $z\sim2$, $z\sim3$, $z\sim5$, $z\sim6$, $z\sim7$, and
  $z\sim8$ samples, consistent with the redshift-dependent fitting
  formula for $\alpha$ derived in Parsa et al.\ (2016).  For context,
  we have overplotted the predictions for completeness vs. shear
  factor from Figure~\ref{fig:predcompshear} for various assumptions
  of source size, using the same color scheme.  Remarkably, we find no
  clear dependence of the relative completeness on the shear factor.
  This suggests that the present sample of extremely faint $\sim
  -15$-mag galaxies have spatial profiles which are indistinguishable
  from that of point sources.\label{fig:compshear}}
\end{figure}

Finally, we should account for the impact that uncertainties in the
magnification and shear maps have on the predicted dependencies
plotted in the left panel of Figure~\ref{fig:predcompshear}.  To
accomplish this, we repeat our quantification of our $z\sim6$
selections as a function of the shear factor but this time using the
median magnification and shear maps created from the seven different
high-resolution lensing models available for the first four HFF
clusters.  The 7 lensing models we consider are the following: CATS
(Jullo \& Kneib 2009; Richard et al.\ 2014; Jauzac et al.\ 2015a,b),
Sharon (Johnson et al.\ 2014), GLAFIC (Oguri 2010; Ishigaki et
al.\ 2015; Kawamata et al.\ 2016), Zitrin-NFW (Zitrin et al.\ 2013,
2015), \textsc{Grale} (Liesenborgs et al.\ 2006; Sebesta et
al.\ 2016), Bradac (2009), and Zitrin-LTM (Zitrin et al.\ 2012, 2015).

The result is shown in the right panel of
Figure~\ref{fig:predcompshear} and contrasted with the dependencies
that only rely on the actual magnification and shear maps.  Not
surprisingly, uncertainties in the magnification models result in a
slight flattening to the predicted completeness vs. shear relation.

\begin{figure*}
\epsscale{0.9}
% /Users/bouwens/magmodel/xdf
\plotone{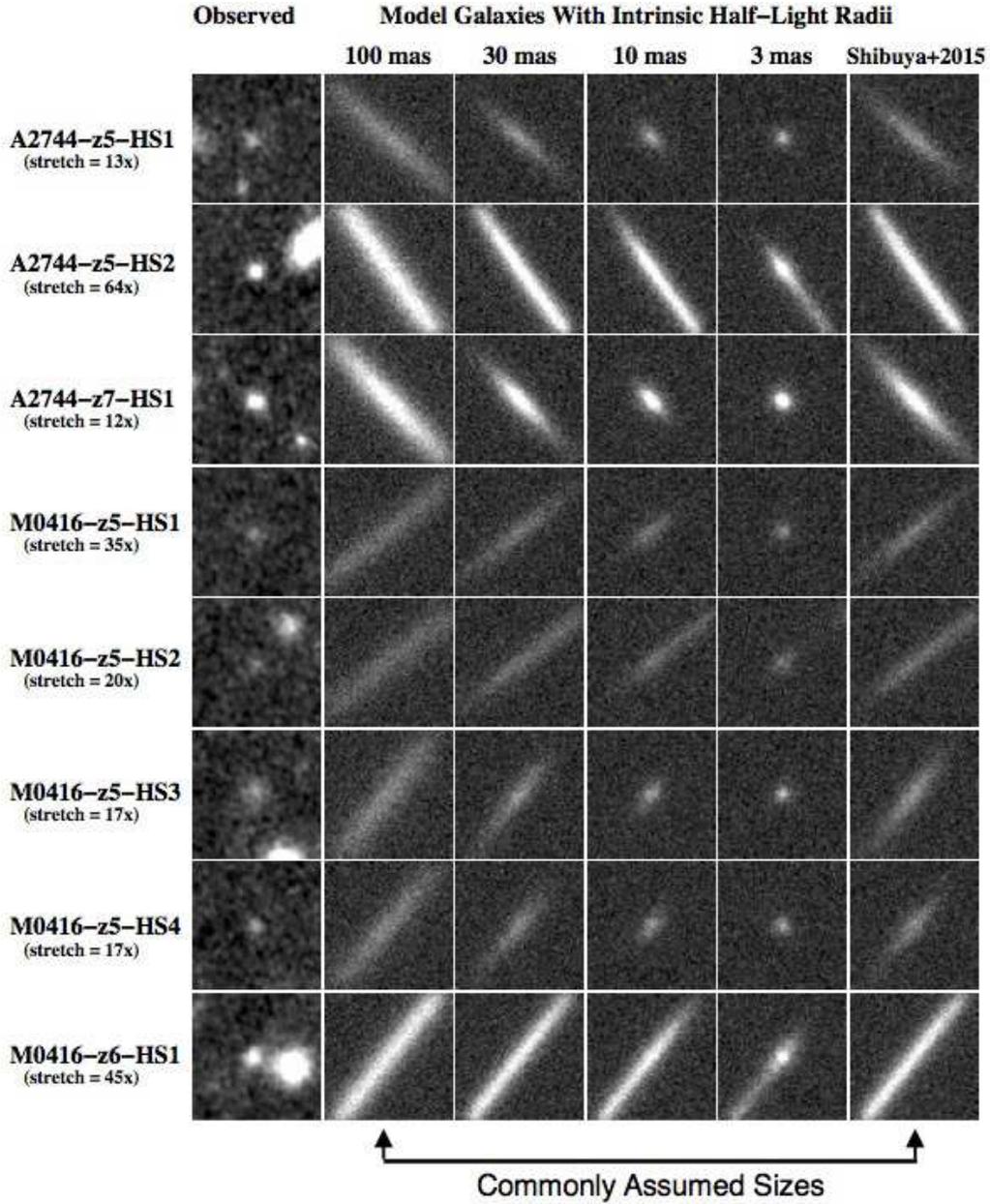}
\caption{Observed vs. predicted images for eight galaxies found behind
  the first 2 HFF clusters.  These galaxies are predicted to have an
  intrinsic magnitude fainter than 30.5 (equivalent to $M_{UV,AB}>-16$
  at $z\sim6$) and a shear factor $>$10.  Indicated for each source is
  the expected magnification factor of each source along the dominant
  shear axis.  The predicted images are realistic, including the
  impact of the {\it HST} WFC3/IR PSF.  We force the model sources to
  have the same flux in a 0.2$''$-arcsec aperture as the observed
  sources, to ensure that the spatial profiles for the model sources
  are clear from this figure.  Shown are the predicted images for the
  sources assuming 4 different values for the half-light radii and the
  Shibuya et al.\ (2015) size-luminosity relation (where sources have
  an intrinsic half-light radius of $\sim$140 mas at $L^*$ and
  $\sim$40 mas at $-$16 mag).  Most previous work assumed a fixed
  half-light radii of $\sim$100 mas or a half-light radius-luminosity
  relation as derived by Shibuya et al.\ (2015).  It is obvious from
  this figure that model sources show much more elongation along the
  shear axis than the observed sources.\label{fig:indprof}}
\end{figure*}

\subsection{Recovered Surface Density vs. Shear: Observations}

Having presented the expectations from our simulations, we now compare
the results with the observations.  To derive this from the
observations, we compute a shear factor for every faint
$H_{160,AB}>28$ source in our $z=5$-8 samples and $H_{160,AB}>26.5$
source in our $z=2$-3 samples whose estimated magnification is $>$10.
Our magnification estimate is taken to be median of the magnification
estimates from the four parametric NFW models, given the photometric
redshift derived for the source.  The shear factor is computed in a
similar way to the magnification factor, using the median of the seven
magnification models presented at the end of \S4.2.

We then bin sources by shear factor and then normalize these totals by
the expected number of sources in each shear-factor bin.  To derive
the expected numbers, we use the fact that the surface density of
sources in a given magnification bin scales as $\phi(L_{lim})dm
\propto (L_{lim}/\mu)^{\alpha+1}/\mu\,dm \propto \mu^{-\alpha-2}
\propto \mu^{-(2+\alpha)}$ where $L_{lim}$ is the limiting luminosity
probed for $z\sim6$ galaxies at $\sim$28.5 mag without the benefit of
gravitational lensing.  The faint-end slope $\alpha$ is taken to have
a value of $-1.45$, $-1.57$, $-1.81$, $-1.93$, $-2.05$, and $-2.17$ at
$z\sim2$, $z\sim3$, $z\sim5$, $z\sim6$, $z\sim7$, and $z\sim8$
samples, consistent with best-fit trend derived by Parsa et
al.\ (2016).  We compute the expected number of sources in each bin by
integrating over the total area in the image plane where sources would
fall in a given ``shear factor'' bin (and where the estimated
magnification would be $>$10) and weighting by the expected surface
densities, i.e., $\mu^{-(2+\alpha)}$.  We then renormalize the
relative completeness so that the average value is equal to 1.  The
results should be directly comparable to the relative completeness
that we just presented in Figure~\ref{fig:predcompshear}.

Figure~\ref{fig:compshear} presents the relative surface densities we
find for galaxies as a function of shear factor for different
groupings of faint sources in redshift, i.e., $z=2$-3, $z=5$, $z=6$-8.
Also included in this same figure are the relative surface density
results combining all of these galaxy samples.

Remarkably, we see no clear dependence of the surface density of faint
$z\sim2$-8 galaxies -- or equivalently the relative completeness -- on
the shear factor, in contrast to the simulation results presented in
Figure~\ref{fig:predcompshear}.  While the statistics are still
modest, one explanation for this result is that if the
highly-magnified sources that we have identified are smaller than
expected.  The implications of this are developed further below where
we establish formally the source sizes that would be consistent with
what is shown in Figure~\ref{fig:compshear}.

Comparing the surface densities of faint galaxies observed in each
shear bin with that predicted making different assumptions about
source size, we find that the results are best matched to the
predictions assuming intrinsic half-light radii of 0, with results
assuming an intrinsic radius of 30 mas and 60 mas being disfavored at
87\% and 99\% confidence.  This assumes that all sources in our
$z=2$-8 samples are of identical size.

In deriving confidence intervals on the sizes of faint sources, we
adopt a flat, uniform prior (0 to 120 mas).  We evaluate the
likelihood of recovering the measured shear factor $S$ distribution
given different assumptions about the true sizes.  We estimate the
likelihood of different source sizes for faint galaxies by creating
several thousand mock realizations of the shear factor $S$
distribution for our faint $z=2$-8 catalogs (with 87 sources each) for
individual size assumptions.  We then ask for which fraction of the
realizations we recover the same number of sources per shear factor
bin $S$ as in the observations to determine the likelihood of a given
size model.  We frame this comparison using the same bins as are
plotted in Figure~\ref{fig:compshear}.  We then sum the total
posterior probability above some assumed source size to determine the
confidence level on the true source size being smaller that value.

A more realistic model to examine is one where galaxies possess a
range in sizes, e.g., as with a log-normal distribution where a
$1\sigma$ scatter of 0.3 dex is assumed (e.g., as in van der Wel et
al.\ 2014; Shibuya et al.\ 2015).  In this scenario, most of the
selected sources in the high-magnification regions are from the small
end of the intrinsic distribution, causing the examined ultra-faint
galaxy population to behave as if it is smaller than it actually is.
If we repeat the above exercise, the observations are again best
matched assuming an intrinsic half-light radius of 0, but where median
half-light radii up to 30 mas and 60 mas are disfavored at 74\% and
99\% confidence, respectively, which is quite a bit larger than if the
size distribution is a delta function.

One could also consider the scenario that many galaxies are near point
sources, but the remainder have much larger sizes.  We take the
intrinsic half-light radii of the smaller mode to be 7.5 mas and the
larger mode to be 60 mas.  We find that up to 36\% of current samples
could be composed of sources with intrinsic half-light radii 60 mas
(assuming a flat prior on the allowed fraction) and still be
consistent with the observed numbers vs. shear factor $S$ at 95\%
confidence.  Given the greater incompleteness of sources with
half-light radii of 60 mas, we estimate that such sources could
compose 79\% of the input samples based on this statistic alone.  For
such a population, this would translate into an incompleteness that is
3.1$\times$ higher than our just assuming near point sources for the
entire ultra-faint population.  This exercise shows the vulnerability
of this test to our assumption that the intrinsic size distribution is
unimodal.  If the size distribution is bimodal, with the fainter mode
essentially 100\% incomplete in our searches, it would have little
impact on our samples or tests we are running.  We keep this caveat in
mind when deriving constraints on the $z\sim6$ LF (Bouwens et
al.\ 2017).

We should consider the possibility that some sources in our sample
could be lower-redshift contaminating sources.  Since we would expect
such contaminating sources to be impacted by gravitational lensing in
a very different way than if they were at the assumed redshifts, we
can reasonably expect their surface densities to be significantly less
dependent on the model magnification factors (or shear factors) for
most of the sources in our selection.  Given this, we would expect
contaminants to show approximately the same surface density in all
shear $S$ bins.  We can estimate the impact of contaminants if we
assume that the non-contaminating sources show the same distribution
as a function of shear factor $S$ as in Figure~\ref{fig:predcompshear}
and also assume that the contaminating sources are identically 1 in
all shear factor $S$ bins.  If we allow for contamination levels of
10\%, which is typical for faint Lyman-break galaxy samples (e.g.,
Bouwens et al.\ 2015; Vulcani et al.\ 2017) and repeat our earlier
estimates assuming identical sizes for all sources, we can find that
we can then exclude the possibility that the half-light radii is 30
mas at 82\% confidence (vs. 87\% confidence in the case of no
contamination).  The likely impact of contamination on our conclusions
is therefore only modest.

On the basis of these tests, it is likely that the typical intrinsic
half-light radii of the ultra-faint ($\sim-15$ mag) galaxies probed
with the HFF program is $<$30 mas (86\% confidence) assuming sources
of the same size and $<$60 mas assuming a log-normal size distribution
(95\% confidence).  This angular size constraint corresponds to
intrinsic half-light radii of $<$165 parsec and $<$330 parsec,
respectively, at $z\sim6$.

Given our application of this test to galaxies with the absolute
magnitude distribution presented in Figure~\ref{fig:hist}, one would
generally expect our size constraints to apply sources in this
luminosity range, i.e., at $\sim-14.7$ mag.  However, if the model
magnification factors were in excess of the true magnification
factors, e.g., as one might expect if errors in the magnification
models were predominately only scattering low magnification sources to
high magnifications, we would underestimate the actual luminosities.
One-sided scatter to high magnification factors could occur at
magnification factors where the lensing models begin to lose their
predictive power, i.e., $\mu\sim20$ (Bouwens et al.\ 2017).  If we
take that factor to be 20, that would suggest a possible underestimate
of the luminosity by $\sim$0.4 mag, which would make the median
absolute magnitude of our sample $\sim-15.1$ mag.

We briefly present the current test in the context of our $z\sim2$-3
samples in \S5.2.

\begin{figure}
\epsscale{0.9}
% /Users/bouwens/magmodel/xdf
\plotone{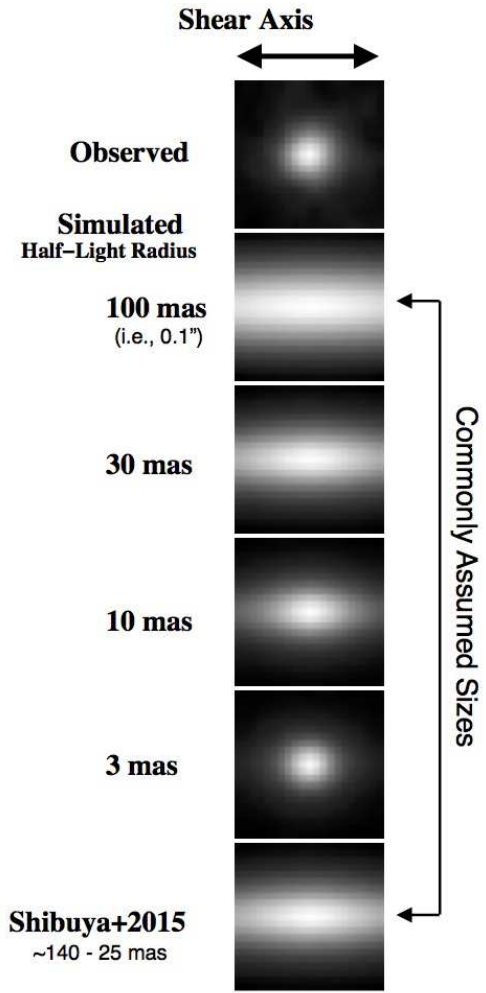}
\caption{Stack of 26 very faint (intrinsic apparent magnitudes
  $>$30.5, equivalent to $M_{UV,AB}>-16$ at $z\sim6$) $z=5$-8 galaxies
  rotated such that the axis predicted to show maximum shear
  elongation lies along the horizontal axis.  Also shown are the
  expected stack results assuming that each of the individual sources
  have intrinsic half-light radii of 100 mas, 30 mas, 10 mas, and 3
  mas, as well as half-light radii dictated by the Shibuya et
  al.\ (2015) size-luminosity relationship.  The simulated stack
  results are very realistic, being constructed from a simulation of
  the $z=5$-8 galaxies behind Abell 2744 and MACS0416 and includes the
  {\it HST} WFC3/IR PSF.  Most previous work assumed a fixed
  half-light radii of $\sim$100 mas or a half-light radius-luminosity
  relation as derived by Shibuya et al.\ (2015).  Amazingly, a stack
  of observed ultra-faint sources only indicates slightly more
  elongation along the shear axis than it does along the axis
  perpendicular to this.  The spatial profile of the stack can be best
  reproduced with intrinsic half-light radii of 4 mas for ultra-faint
  $z=5$-8 galaxies.\label{fig:stackprof}}
\end{figure}

\begin{figure*}
\epsscale{1.15}
\plotone{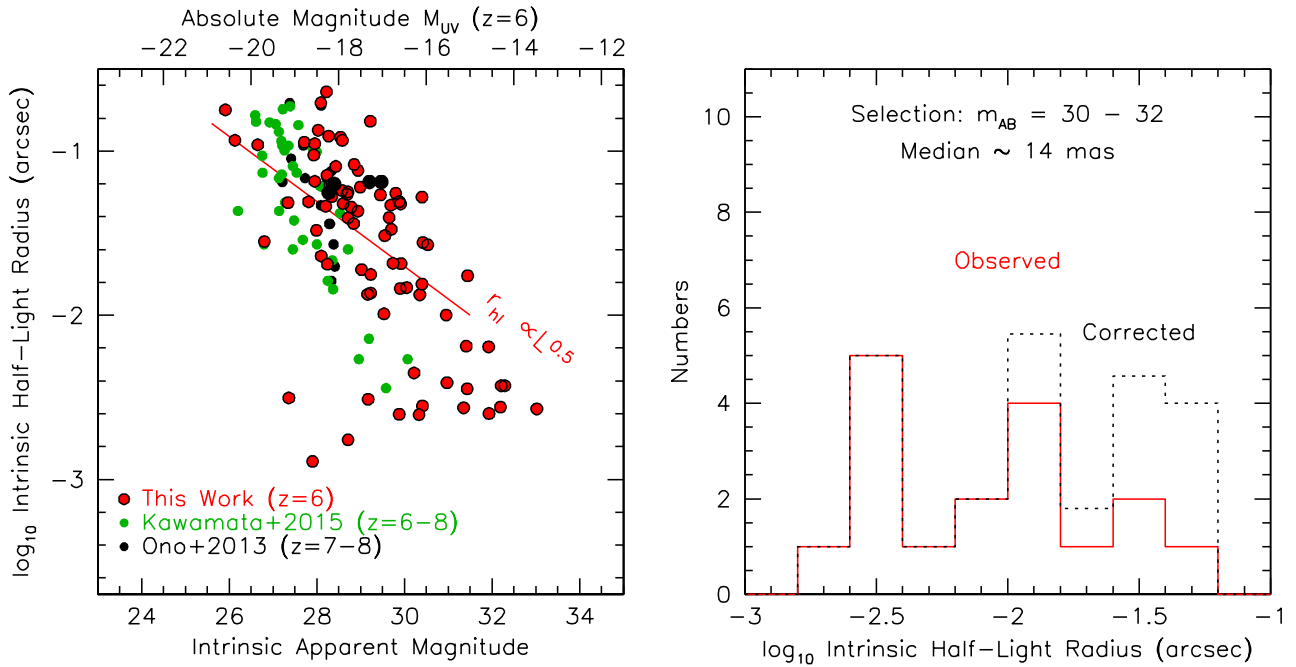}
\caption{(\textit{left}) Inferred half-light radii vs. intrinsic
  apparent magnitude for $z\sim6$ sources behind Abell 2744 and
  MACS0416 based on fits with \textsc{galfit} and the CATS lensing
  models.  For context, we also show the size measurements for $z=7$-8
  galaxies from the HUDF (Ono et al.\ 2013: \textit{black circles})
  and from $z=6$-8 galaxies inferred from the Abell 2744 cluster and
  parallel field by Kawamata et al.\ (2015: \textit{green circles}).
  The solid red lines shows one possible size-luminosity relation that
  seems consistent with the observational constraints from this study
  (where $r_{hl} \propto L^{0.50}$).  (\textit{right}) Number of
  intrinsically-faint $z\sim6$ galaxies (with intrinsic $H_{160,AB}$
  magnitudes 30-32 mag) vs. inferred intrinsic half-light radius.  The
  dotted histogram incorporates a correction for incompleteness, to
  account for the higher efficiency of selecting sources with
  intrinsic half-light radii $r_{hl}\sim$7.5 mas in the magnitude
  range $H_{160,AB}=30$-32 than with intrinsic radii $r_{hl}>10$ mas.
  The median intrinsic half-light radius in the magnitude interval
  30-32 is inferred to be 14$_{-3}^{+5}$ mas, slightly larger than
  inferred elsewhere in the analysis due to a few larger sources with
  intrinsic magnitudes 30.0-30.5 mag.\label{fig:galfit}}
\end{figure*}

\section{Direct Size Measurements on Individual / Stacked Sources}

Results from the previous section provided us with strong evidence
that the half-light radii of extremely faint ($\sim$$-15$ mag)
$z\sim2$-8 galaxies is $<$30 mas for sources of identical size and
$<$60 mas if we consider galaxies to show a range of sizes.  30 mas
and 60 mas correspond to 165 pc and 330 pc, respectively, at $z\sim6$.

The purpose of the present section is to try to confirm these
constraints through direct measurements of the source sizes for faint
galaxies.  Despite the clear value of such measurements, it is
important to remember that we can only obtain them for selected
sources.  It is possible that more extended sources could exist but
fail to be selected.  This is why the shear-based technique from the
previous section is useful, since it provides us with a method to
infer the size distribution even in the presence of incompleteness.

We focus our attention on those sources which are expected to
stretched by a factor of $>$10 along one axis, so that we can probe
the sizes of faint galaxies to a resolution of $<$10 mas.  In
determining which sources we might expect to exhibit such spatial
elongation, we expressly make use of the median lensing maps, since we
will make use of the latest deflection maps from the CATS team in
modeling the spatial structure of sources.  By using different lensing
maps for the selection and measurement steps, we ensure that the
present results are more robust against uncertainties in the lensing
maps.

As in \S4, we examine sources in the faintest magnitude bin, since
this provides us with the most leverage in probing the faintest
galaxies observed in the HFFs.  We restrict our attention to results
over the first two HFF clusters given the greater maturity of the
available magnification models.

We focus first on extremely faint galaxies in the general redshift
range $z\sim5$-8 and then move onto faint galaxies in the redshift
range $z\sim2$-3.

\subsection{$z=5$-8 Samples}

We consider first faint galaixes in the redshift range $z\sim5$-8.  We
find 26 such faint galaxies behind those two clusters that are
stretched by more than a factor of 10 along a single axis and where
the intrinsic apparent magnitude is $>$30.5 mag (corresponding to
$>-16$ mag at $z\sim6$).  This list of 26 sources includes 17
$z\sim5$, 5 $z\sim6$, 1 $z\sim7$, and 3 $z\sim8$ galaxies from our
sample.  The median $M_{UV}$ magnitude for these sources is
$\sim$$-$15 mag, very similar to the sample used in the previous
section.

We present postage stamps of eight randomly-chosen sources from this
list in Figure~\ref{fig:indprof} and contrast their spatial profiles
with that predicted for circular sources based on current lensing
models.  {\it It is striking to see in Figure~\ref{fig:indprof} that
  model sources show a remarkable degree of elongation along the shear
  axis compared to the observed sources.}  Amazingly, we find this to
be the case, even assuming source sizes of $r_{hl}=10$ mas.  Similar
to the results from the previous section, this suggests that faint
galaxies are extremely small.\footnote{We remark that our use of
  sources with circular profiles (instead with non-unity axis ratios)
  does not fundamentally change this result.  We would simply expect
  non-circular sources to appear larger or smaller depending on
  whether their major axis lie along the axis of maximum shear or
  not.}

We can obtain a higher S/N look at the spatial profile of faint
galaxies by taking the 26 sources from these samples, rotating the
images of the sources so that they lie along the horizontal axis, and
then combining the images to create a deep stack.  We perform the same
exercise on model images of the cluster where we create these images
by applying a deflection map of sources behind a cluster to a bunch of
model sources with fixed intrinsic size as well as adopting the
Shibuya et al.\ (2015) half-light radius size scaling.  The stack
results are presented in Figure~\ref{fig:stackprof}.  Fitting the
stack result with \textsc{galfit}, we measure a half-light radius of
64$\pm$1 mas along the predicted axis of maximum shear.  After
correction for the impact of lensing (the median estimated elongation
along the shear axis is 17 for sources contributing to the stack), the
measured size translates to an intrinsic half-light radius of 4 mas.

One potential concern about the probe of source size featured i n
Figure~\ref{fig:indprof}-\ref{fig:stackprof} is the possibility that
the featured sources are interpreted to be stretched by much larger
factors along the major shear axis than in reality, due to errors in
the lensing models.  One can investigate the impact of such errors on
analyses like that featured here, by looking at how well the median
stretch factor $S^{1/2} \mu^{1/2}$ predicts the stretch factor from
one of the parametric models (CATS, GLAFIC, Zitrin-NFW, Sharon) in the
median.  In Bouwens et al. (2017, in prep), we show that the median
$S^{1/2} \mu^{1/2}$ map is predictive to factors of 10, with symmetric
scatter about that value, but higher than that, the dominant scatter
is in the direction from high to low values.  This suggests that we
may systematically overestimate the magnification along the major
shear axis by a factor of $\sim$2 for the typical source and so
constraints on the measured size may be closer to 8 mas, instead of 4
mas.

\begin{deluxetable*}{ccccccccc}
\tablewidth{0pt}
\tablecolumns{11}
\tabletypesize{\footnotesize}
\tablecaption{Coordinates and other measured properties of $z\sim2$-8
  sources used in our analysis.\tablenotemark{*}\label{tab:catalog}}
\tablehead{
\colhead{ID} & \colhead{R.A.} & \colhead{Dec} & \colhead{$m_{AB}$} & \colhead{$M_{AB}$} & \colhead{$z_{phot}$} & \colhead{$\mu$\tablenotemark{a}} & \colhead{$S$\tablenotemark{b}} & \colhead{$r_{hl}$\tablenotemark{c}}}
\startdata
A2744I-4242524441 & 00:14:24.257 & $-$30:24:44.11 & 28.03 & $-$16.23 & 6.10 & 9.20 & 6.90 & 0.141$\pm$0.010 \\
A2744I-4231724324 & 00:14:23.172 & $-$30:24:32.44 & 27.19 & $-$17.74 & 5.62 & 4.98 & 3.25 & 0.293$\pm$0.010 \\
A2744I-4226324225 & 00:14:22.639 & $-$30:24:22.51 & 28.09 & $-$17.06 & 5.62 & 4.04 & 1.92 & 0.069$\pm$0.330 \\
A2744I-4223024479 & 00:14:22.306 & $-$30:24:47.98 & 27.48 & $-$16.67 & 5.96 & 10.18 & 6.21 & 0.071$\pm$0.103 \\
A2744I-4197224471 & 00:14:19.728 & $-$30:24:47.10 & 28.29 & $-$17.33 & 5.96 & 2.64 & 2.03 & 0.247$\pm$0.030 \\
A2744I-4219124454 & 00:14:21.910 & $-$30:24:45.46 & 29.24 & $-$15.02 & 6.10 & 9.17 & 6.19 & 0.024$\pm$0.010 \\
A2744I-4169524527 & 00:14:16.956 & $-$30:24:52.79 & 26.10 & $-$20.05 & 6.10 & 1.62 & 1.48 & 0.139$\pm$0.040 \\
A2744I-4169624404 & 00:14:16.960 & $-$30:24:40.40 & 28.08 & $-$17.90 & 5.62 & 1.89 & 1.61 & 0.407$\pm$0.020
\enddata
\tablenotetext{*}{Table~\ref{tab:catalog} is published in its entirety
  in the electronic edition of the Astrophysical Journal.  A portion
  is shown here for guidance regarding its form and content.}
\tablenotetext{a}{Magnification factor adopted in this analysis.  Median of the four parametric models.}
\tablenotetext{b}{Shear factor $S$ adopted in this analysis.  Median of the available models.}
\tablenotetext{c}{Circularized half-light radii in kpc.}
\end{deluxetable*}

The intrinsic half-light radius inferred for faint sources is
sufficiently small that it is useful to try a similar test on more
luminous sources where the size distribution is more well established
from various studies in the literature.  We consider such a test in
Appendix A looking at all sources in the magnitude range 27.0 to 29.4
mag, which are stretched by at least a factor of 6 along a single
axis.  The results of this are shown in Figure~\ref{fig:indprofzb} of
the Appendix A.  In contrast to the results of this section, the
observed sizes of the moderate luminosity sources are in reasonable
agreement with the expected sizes using the Shibuya et al.\ (2015)
relations as a guide.

As a final check on these results, it is useful to consider direct
size measurements with \textsc{galfit} on the full set of $z\sim6$
galaxies behind the same two HFF clusters just considered.  For these
fits, we take the Sersic index to equal 1, coadd the $Y_{105}$,
$J_{125}$, $JH_{140}$, and $H_{160}$ images together weighting by the
inverse variance, and then fit the profiles, taking as the PSF a
similar inverse variance weighting of the $Y_{105}$, $J_{125}$,
$JH_{140}$, and $H_{160}$ PSFs.  We take the measured intrinsic
half-light radius to be equal to the measured half-light radius,
divided by magnification factor along the major shear axis, i.e,
$\mu^{1/2} S^{1/2}$ where $S$ is the shear factor.  We treat sources
as being circular for these corrections given that lack of correlation
between the observed and predicted major axis for most sources.  The
results are shown in the left panel of Figure~\ref{fig:galfit}, also
including the size measurements from Ono et al.\ (2013) using the
HUDF09 and the HUDF12 data (Bouwens et al.\ 2011; Ellis et al.\ 2013;
Koekemoer et al.\ 2013) and from Kawamata et al.\ (2015) using the HFF
cluster and parallel data over Abell 2744.  Typical uncertainties in
the measured half-light radii is 20 mas, equivalent to $\sim$0.1 kpc
(prior to incorporating the additional size leverage provided by the
lensing magnification).  Size measurements for the sources are
provided in Table~\ref{tab:catalog}.

In the right panel of Figure~\ref{fig:galfit}, we present a histogram
of the inferred sizes for sources inferred to have the faintest
apparent magnitude in these fields.  For sources with intrinsic
half-light radii $>$10 mas, we also show a dotted histogram to
indicate the expected number of sources we would find, if we had
included a correction for the incompleteness of these sources.  We can
estimate the incompleteness using the same simulations described in
\S4.  The median half-light radius that we infer for intrinsic faint
30-32 mag galaxies is 14$_{-3}^{+5}$ mas ($\sim$80 pc at $z\sim6$).

We include on Figure~\ref{fig:galfit} one possible size-luminosity
relation, i.e., $r_{hl}\propto L_{UV} ^{0.5}$, that appears consistent
with most of the brighter constraints from the current study as well
as the literature.  Despite the indicative fit, the exponent to the
size-luminosity relation is fairly uncertain.  One can derive a
conservative lower limit to the uncertainty based on the estimated
error in the median intrinsic half-light radius from the right panel.
Based on that estimate and a luminosity baseline of 5.5 mag from $L^*$
(well constrained by bright samples: e.g., Shibuya et al.\ 2015) to
$-$15.5 mag (our faint sample), we estimate an uncertainty of
$\pm$0.07 in the exponent.  However, we caution that the true error
could be larger.\footnote{One potential point of concern is the fact
  that the exponent 0.5 we derive for the size-luminosity relation is
  the value one would expect, if surface brightness selection biases
  dominated the sample composition.}

\subsection{$z=2$-3 Samples}

Secondly, we focus on size measurement for extremely faint galaxies at
$z\sim2$-3.  As in \S5.1, we focus on those sources which are
especially magnified, i.e., $>$10$\times$, along one of the two axes,
and where the inferred intrinsic apparent magnitudes.  There are six
such magnified sources in our $z\sim2$-3 samples, again only making
use of those clusters with the most mature magnification models
(MACS0416 and Abell 2744).  As we only have $z\sim2$-3 samples over
Abell 2744, the six sources are drawn from data over that cluster.

\begin{figure*}
\epsscale{0.9}
% /Users/bouwens/magmodel/xdf
\plotone{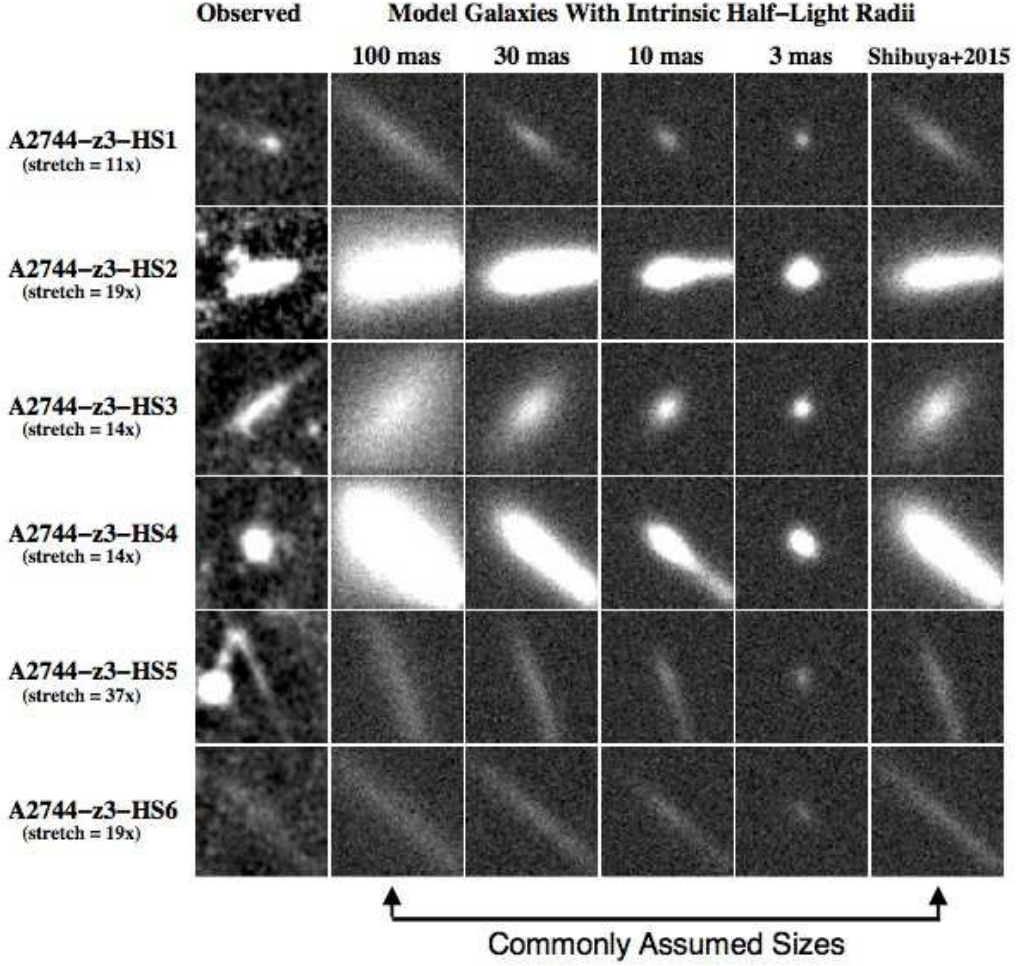}
\caption{Similar to Figure~\ref{fig:indprof} but for galaxies from our
  $z\sim3$ samples available over Abell 2744 with the most refined
  magnification models.  The intrinsic apparent magnitudes for the
  plotted sources are $>$29 mag, equivalent to $>-$16.5 mag, according
  to the median magnification model.  Most previous work assumed a
  fixed half-light radii of $\sim$100 mas or a half-light
  radius-luminosity relation as derived by Shibuya et al.\ (2015).  It
  is obvious from this figure that the observed sources show less
  elongation along the shear axis than for the largest size
  assumptions, being most consistent with intrinsic sizes of 3 mas, 10
  mas, and 30 mas for the observed sources.\label{fig:indprofz3}}
\end{figure*}

Results for individual sources are presented in
Figure~\ref{fig:indprofz3}.  In contrast to results presented for our
$z\sim5$-8 samples, extremely faint $z\sim2$-3 sources do show
evidence for being moderately extended along the expected shear axis.
An intrinsic half-light radius of $\sim$10 mas provides a reasonable
representation of A2744-z3-HS1 and A2744-z3-HS2, while $\sim$30 mas
provides a reasonable representation for A2744-z3-HS3, A2744-z3-HS5,
and A2744-z3-HS6.  A2744-z3-HS4 is most consistent with an intrinsic
half-light radius of 3 mas.

\begin{figure}
\epsscale{0.8}
% /Users/bouwens/magmodel/xdf
\plotone{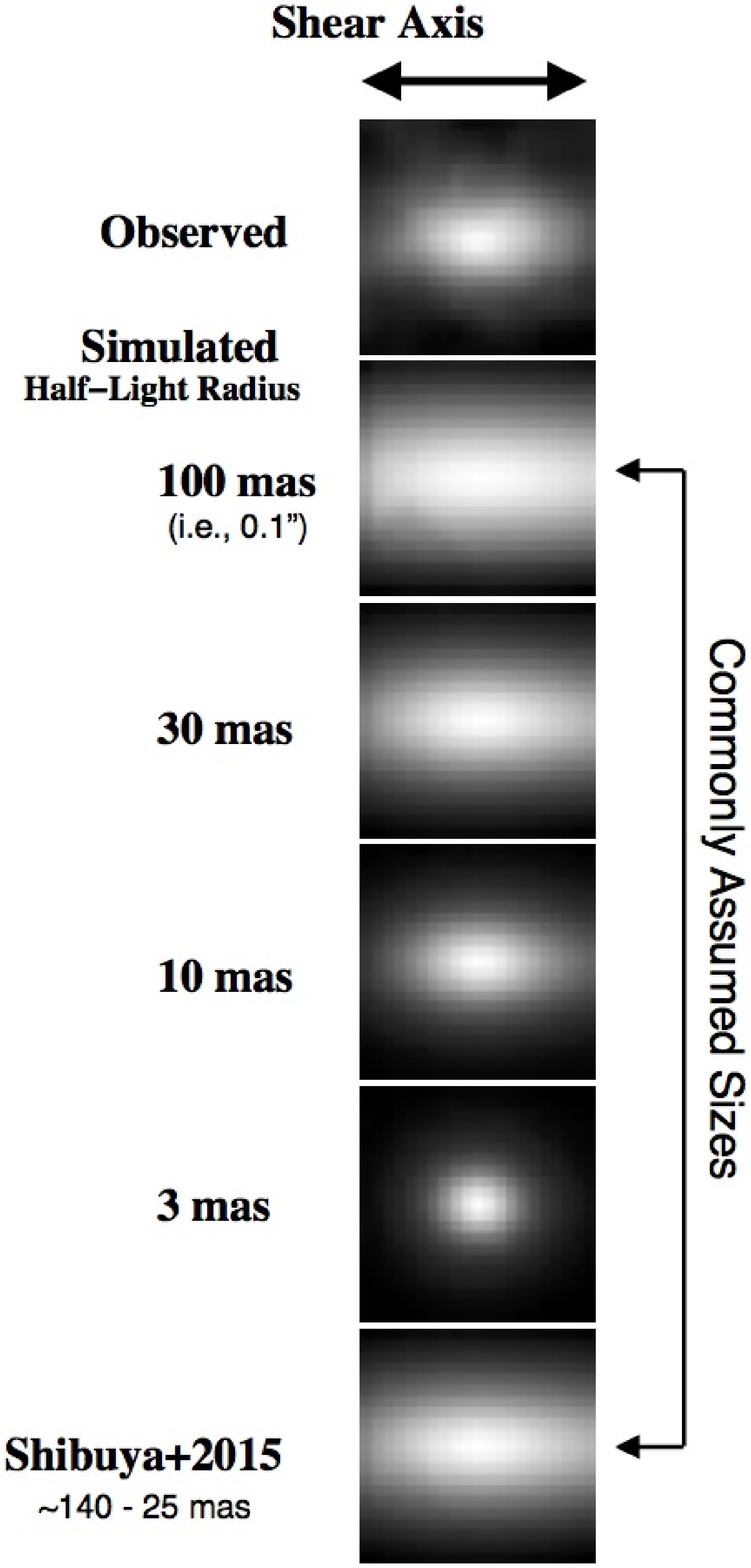}
\caption{Similar to Figure~\ref{fig:stackprof} but for very faint
  (intrinsic apparent magnitudes $>$29, equivalent to
  $M_{UV,AB}>-16.5$) $z\sim3$ galaxies.  A stack of observed
  ultra-faint sources indicates slightly more elongation along the
  shear axis than it does along the axis perpendicular to this.  The
  spatial profile of the stack can be best reproduced with intrinsic
  half-light radii of 11 mas for ultra-faint $z\sim3$
  galaxies.\label{fig:stackprofz3}}
\end{figure}

Rotating sources such that their major shear axes are aligned and
stacking, we compare the stacked image with that expected for
different intrinsic size models in Figure~\ref{fig:stackprofz3}.  In
contrast again to the results for our $z\sim5$-8 sample, we find that
the image stack shows extension along the expected shear axis.  The
stacked profile agrees best with the results assuming intrinsic
half-light radii of 10 mas.

Results from this subsection again are consistent with very small
sizes for faint galaxies in the redshift range $z\sim2$-3, but are
suggestive of somewhat larger sizes for sources than in the range
$z\sim5$-8.  Sources in Figure~\ref{fig:indprofz3} have inferred
physical sizes ranging from $\sim$20 pc (3 mas) to $\sim$250 pc (30
mas), while the stack has an intrinsic half-light radius consistent
with $\sim$80 pc (11 mas: Figure~\ref{fig:stackprofz3}).

Considered by itself, the relative completeness results for the
$z\sim2$-3 sample presented in \S4 also are suggestive of larger sizes
for the $z\sim2$-3 sample, given the higher values of the completeness
at low shear factors and lower values at high shear factors.  The
observed trend presented in Figure~\ref{fig:compshear} agrees best
with the 15-mas model, but the numbers in the $z\sim2$-3 sample are
sufficiently low that we cannot rule out 34-mas and 60-mas sizes for
faint $z=2$-3 sources at 68\% and 91\% confidence, respectively.

\section{Discussion}

The results of the tests we have performed in the previous two
sections strongly suggest that the faintest galaxies accessible from
the HFF program are very small, with probable intrinsic half-light
radii of $<$165 pc at $z\sim6$ and $<$240 pc at $z\sim2$-3.  Direct
fits to the sizes of many individual sources result in much smaller
sizes, i.e., from 3 mas to 14 mas.  Table~\ref{tab:summary} provides a
summary of the different tests we have performed to try to constrain
the size distribution in faint $z=2$-8 galaxies.

\subsection{Comparison with Previous Results}

Before interpreting the implications of the present results on the
sizes of extremely faint galaxies over the HFF, it is useful to
compare with previous work on the sizes of galaxies, as inferred from
the HUDF, CANDELS, and the first HFF cluster and parallel field.

The most comprehensive recent work on the sizes of star-forming
galaxies in the distant universe is by Shibuya et al.\ (2015), who
looked at 180,000 individual sources found over the CANDELS fields,
the first 2 HFF clusters and parallels, and the HUDF and
systematically quantified the size distribution of galaxies as a
function of redshift, luminosity, stellar mass, and also rest-frame
$UV$ color.  Shibuya et al.\ (2015) found that the half-light radius
of $L^*$ galaxies in the distant universe is approximately $\sim$1
kpc, with $r_{hl}$ correlating with luminosity $L$ as $r\propto L_{UV}
^{0.27}$, such that sources with absolute magnitudes of $-18$, $-16$,
and $-14$ mag would have sizes of 0.38 kpc, 0.23 kpc, and 0.14 kpc,
respectively (assuming $r_{hl}\sim0.8$ kpc at $-21$ mag as indicated
by their Figure 10).  

The observed sizes of galaxies from other recent studies (e.g. Huang
et al.\ 2013) are comparable to what was found by Shibuya et
al.\ (2015).  Ono et al.\ (2013) reported stacked sizes of $\sim$0.3
to 0.35 kpc for $z\sim7$-8 galaxies found in the HUDF at $\sim$28.2
mag and $\sim$29.2 mag.  The measured half-light radii of $z\sim6$-8
sources from Kawamata et al.\ (2015) occupy a similar locus in the
half-light radii vs. luminosity plane as found by Ono et al.\ (2013).
In general these sizes are $\sim$3-4$\times$ larger than we find using
the present constraints.

Intriguingly though, Kawamata et al. (2015) find a few lensed galaxies
over the first HFF cluster field with inferred sizes very similar to
what we find here, e.g., $\sim$30-50 pc.  Given the presence of
surface brightness selection effects against larger, lower surface
brightness galaxies, Kawamata et al.\ (2015) could not know whether
the small sizes they measured for galaxies in their faintest
luminosity subsample, i.e., $\sim$30-100 pc, were representative or
not.  The present results suggest that such small sources are indeed
ubiquitous in faint samples of $z=2$-8 galaxies, with many sources in
our samples having apparent sizes of $\sim$3-10 mas (17-55 pc at
$z\sim6$).  Moreover, indicative fits to the size - luminosity
relation yield a steeper dependence on luminosity than derived by
Shibuya et al.\ (2015) based on more luminous samples.

It is unclear why the sizes of extremely faint galaxies might differ
so dramatically from what is found at the bright end of the UV LF.
Interesting, these new results could be revealing that the lowest
luminosity galaxies are really dominated at any time by one or two
localized regions of star formation.  These star forming regions are
also striking in that they appear to be so compact, with many having
apparent sizes consistent with just 20-50 pc.  This is similar or
smaller than the sizes (10-100 pc) of many giant molecular clouds
(GMCs) and also most of the star-forming clumps seen in the local
universe by SINGS (Kennicutt et al.\ 2003) or in lensed galaxies at
$z=1$-4 (Livermore et al.\ 2012, 2015).

\begin{deluxetable}{ccccccc}
\tablewidth{0cm} \tablecolumns{7} \tabletypesize{\footnotesize}
\tablecaption{Constraints on the Sizes of Extremely Faint
  ($H_{160,AB}>30.5$) $z=2$-8 Galaxies seen behind the HFF
  clusters\label{tab:summary}} \tablehead{\colhead{Description} &
  \colhead{Constraint}} \startdata \multicolumn{2}{c}{Completeness
  vs. Shear Factor (\S4)}\\ Assuming Identical Sizes & $<$30 mas
(87\%) \\ & $<$60 mas (99\%) \\ Log-Normal, Width=0.3dex & $<$30 mas
(74\%) \\ & $<$60 mas (99\%) \\\\

\multicolumn{2}{c}{Direct Fits (\S5: \textsc{galfit})\tablenotemark{a}}\\
Individual Sizes (Figure~\ref{fig:indprof}) & $\lesssim$3 mas \\
Size of Stack (Figure~\ref{fig:stackprof}) & 4 mas \\
Size Distribution (Figure~\ref{fig:assump}) & 14$_{-3}^{+5}$ mas \\
($H_{160,AB}=30$-32)
\enddata
\tablenotetext{a}{These size measures would include the selected sources, and hence could potentially miss larger, lower-surface brightness galaxies}

\end{deluxetable}

\begin{figure}
\epsscale{1.15}
\plotone{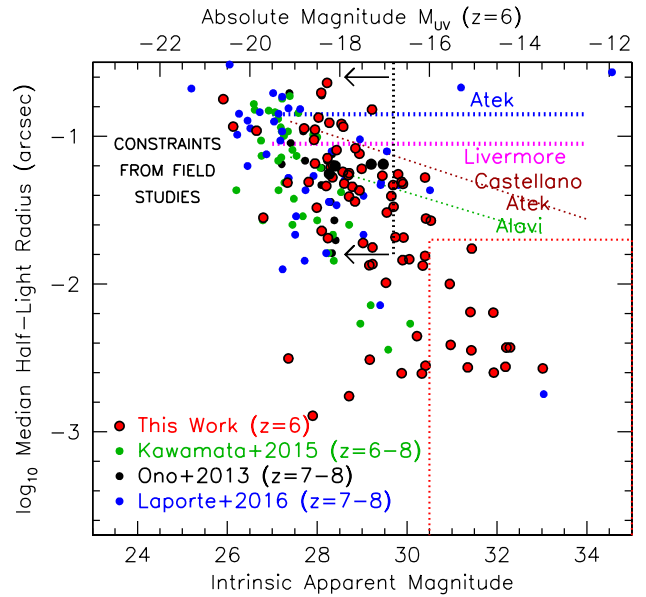}
\caption{Comparison of various observational constraints on the
  measured sizes of faint $z=6$-8 galaxies from the HUDF (Ono et
  al.\ 2013: \textit{black circles}), the Abell 2744 HFF cluster +
  parallel field (Kawamata et al.\ 2015: \textit{green circles}), the
  MACS0416 + MACS0717 HFF cluster + parallel field (Laporte et
  al.\ 2016: \textit{blue circles}), as well as the measured and
  magnification-corrected sizes of sources behind Abell 2744 and
  MACS0416 (\textit{red circles}) using \textsc{galfit}.  The dotted
  red box shows the constraints we obtain via our indirect arguments.
  These constraints are shown in relation to the assumptions that have
  been made in a number of recent studies looking at the $z=2$-8 $UV$
  LFs (\textit{indicated with solid + dotted lines}).  In particular,
  Alavi et al.\ (2016) and Castellano et al.\ (2016b) had assumed that
  the faint galaxies had sizes governed by an extrapolation of the
  Shibuya et al.\ (2015) and Huang et al.\ (2013) size-luminosity
  relation to $>$29 mag, and Livermore et al.\ (2017) assumed median
  sizes of 0.5 kpc (equivalent to an intrinsic half-light radius of
  $\sim$90 mas).  Atek et al.\ (2014, 2015) assumed both intrinsic
  half-light radii of $\sim$150 mas and that galaxies follow the Huang
  et al.\ (2013) size-luminosity scalings.  Each of these assumptions
  was plausible, as each reproduced the sizes of galaxies towards the
  faint end of the HUDF observations.  However, the present
  observations suggest that faint galaxies have even smaller sizes
  than what had been assumed in most previous work.  Even if the
  smaller sizes that we find here are a significant fraction of the
  sample at $>$29 mag, the corrections rapidly diverge for larger
  sizes and so any sample at such magnitudes is going to rapidly
  become uncorrectable for all practical purposes, leading to quite
  inaccurate volume densities and LFs.\label{fig:assump}}
\end{figure}

\begin{figure}
%/Dropbox/lensed_samples/gencat/livermore
\epsscale{1.03}
\plotone{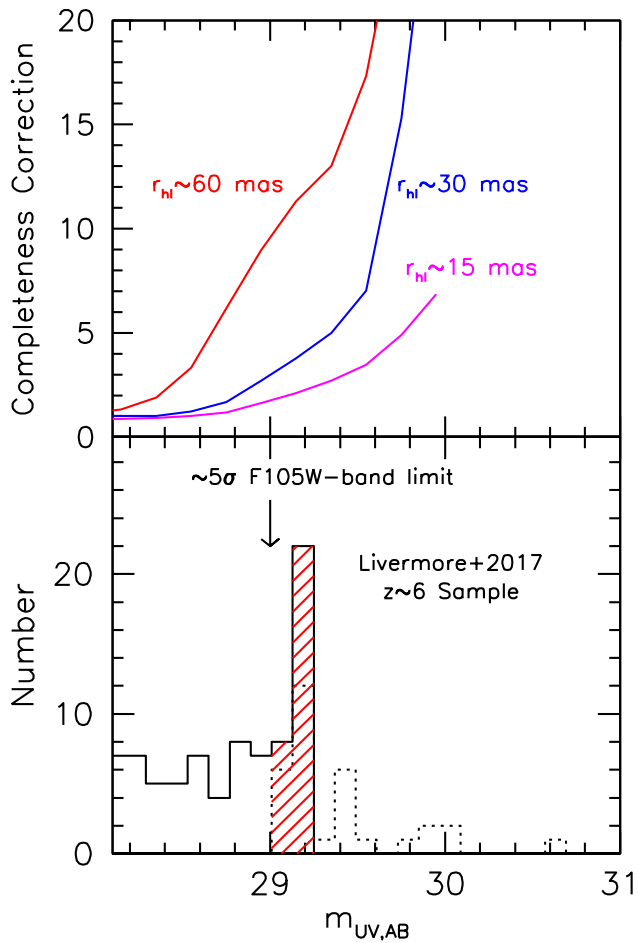}
\caption{Illustration of the impact that assumptions about source size
  can have on the volume density inferred for $z\sim6$ galaxies at the
  faint end of one's selection.  (\textit{upper}) Approximate
  completeness corrections for a $z\sim6$ selection vs. apparent
  magnitude derived from simulations assuming intrinsic half-light
  radii of 60 mas, 30 mas, and 15 mas for sources behind lensing
  clusters and a magnification factor of 5.  Clearly, the estimated
  selection volume at $>$29 mag is very sensitive to assumptions about
  source size, as even small changes can have a large impact.
  (\textit{lower}) \# of $z\sim6$ sources vs. apparent magnitude from
  Livermore et al.\ (2017), with the faint $>$29 mag sources indicated
  with a hatched red shading.  The apparent magnitudes presented here
  for individual sources are derived based on the magnifications,
  absolute magnitudes, and redshifts given in Table 7 of Livermore et
  al.\ (2017).  The inclusion of such faint sources can introduce
  large biases if assumptions about the source size are not correct.
  This is particularly problematic when the observed counts show such
  a large excess of sources at the completeness limit, as is the case
  for the Livermore et al.\ (2017) sample with 22 sources in the
  $m_{AB}\sim29.2$ bin ($5\sigma$ higher than adjacent bins).  All 22
  sources in this bin had measured magnitudes of $>$29.13-mag in the
  earlier Livermore et al.\ 2016 catalogs [\textit{dotted
      histogram}]).  This large pile-up of sources at the $z\sim 6$
  magnitude limit is not apparent in Figure 9 of L17, since L17 set
  the upper vertical axis to 30 -- even though there are actually 45
  sources in their faintest bin.\label{fig:liver1}}
\end{figure}

\begin{figure*}
%/Dropbox/lensed_samples/gencat/livermore
\epsscale{0.85}
\plotone{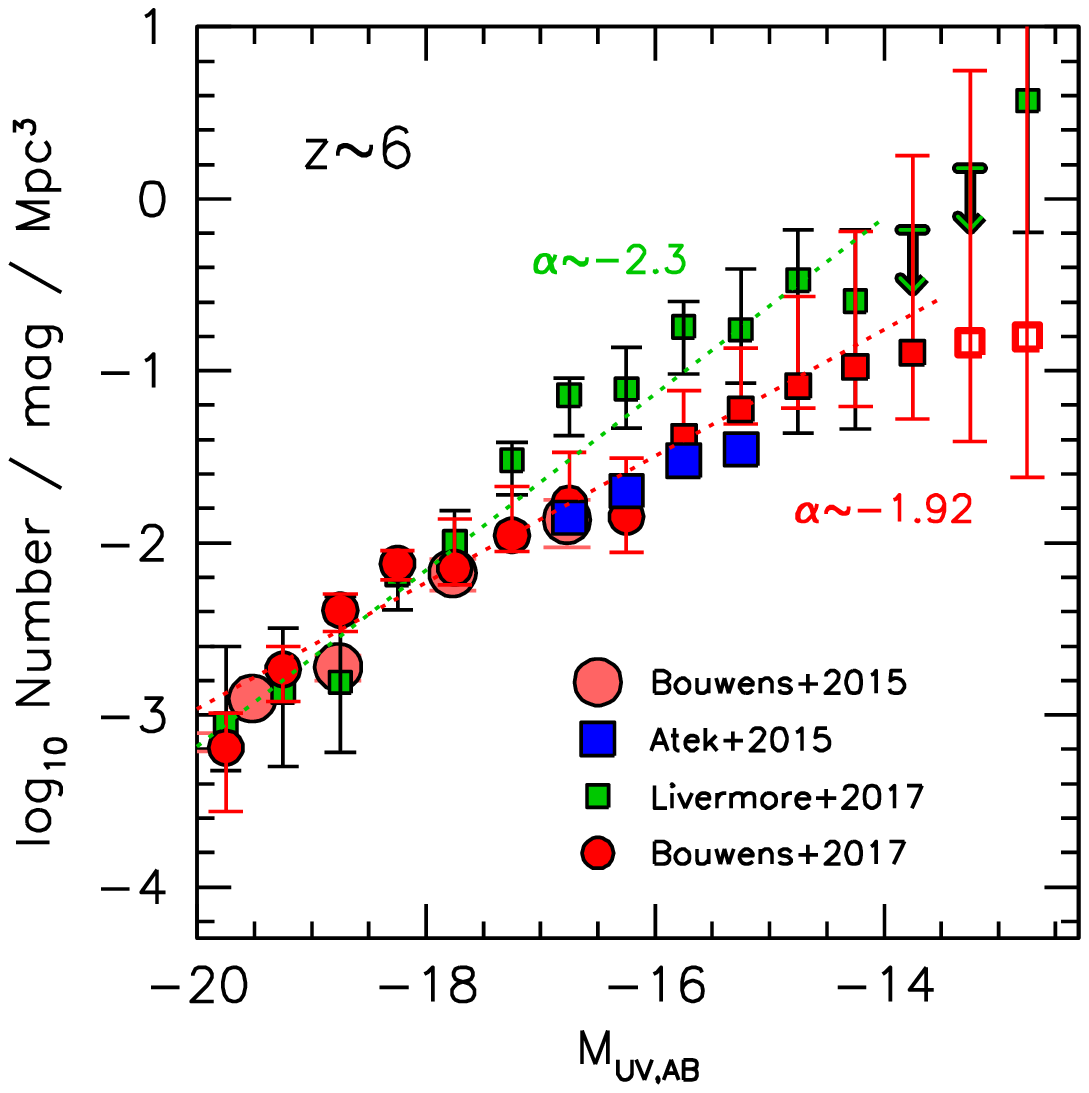}
\caption{Rest-frame $UV$ LF at $z\sim6$, as estimated by Atek et
  al.\ (2015: \textit{blue squares}), Livermore et al.\ (2017:
  \textit{green squares} [extracted from their Figure 11]), and
  Bouwens et al.\ (2017: \textit{red circles}).  Faintward of $-16$
  mag, the results of Bouwens et al.\ (2017) are given by the best-fit
  model from that study with the upper and lower error bars giving the
  68\% confidence region.  Sources in the Bouwens et al.\ (2017)
  faintward of $-13.5$ mag are shown with open squares to indicate
  their greater uncertainties.  For context, the $z\sim6$ $UV$ LF from
  Bouwens et al.\ (2015) based on fields like the HUDF + CANDELS is
  also shown (\textit{light red circles}).  Also shown is the
  faint-end slope of $\sim-2.3$ that is a good representation to the
  uncorrected Livermore et al.\ (2017) LF results (assuming a minimal
  Eddington bias correction that our simulations suggest is likely
  appropriate) and $\sim-1.92$ as implied by the Bouwens et
  al.\ (2017) results using the first four HFF clusters.  For
  consistency in the luminosities presented, the luminosities of the
  individual points in the Livermore et al.\ (2017) LF have been
  corrected brightward by $\sim$0.25 mag to account for a similar
  offset between their measured apparent magnitudes and those from our
  own study (see Bouwens et al.\ 2017).  Higher volume densities are
  reported by Livermore et al.\ (2017) at $z\sim6$ relative to the
  Atek et al.\ (2015) and Bouwens et al.\ (2017) results.  This is
  likely the result of their inclusion of many $>$29-mag sources and
  their assumption of larger sizes for the $>$29-mag sources than we
  now find.  See \S6.2 and
  Figure~\ref{fig:lf6size}.\label{fig:liver2}}
\end{figure*}

\subsection{Impact on the Completeness of Faint Samples}

The small source sizes for galaxies inferred here have significant
implications on the derived completeness of faint galaxies, and thus
on the derivation of the faint-end slope of the UV LFs.  Assumptions
made about the size distribution can have a dominant impact on the
derived faint-end slope (see Figure~\ref{fig:lf6size}).

The present findings imply that faint $z=2$-3 + $z=5$-8 galaxies would
be much easier to select than has assumed to be the case in many
previous studies, where almost universally larger source sizes for
extremely faint galaxies have been assumed.  Source sizes from
$r_{hl}\sim40$-130 mas (Castellano et al.\ 2016b) to $\sim$150 mas
have been assumed (Atek et al.\ 2015).  In Figure~\ref{fig:assump}, we
indicate the assumptions that have been utilized in different studies
and contrast those assumptions with what we have found here and with
the measured sizes of many sources in the literature (Ono et
al.\ 2013; Kawamata et al.\ 2015; Laporte et al.\ 2016).

Inspecting the results of Alavi et al.\ (2016), we see that their
estimated completeness is 80-90\% and $\sim$30\% at $\sim$28 and
$\sim$29 mag, respectively, while for the Atek et al.\ (2015) studies
the approximate completeness is 60\% and 10\%, respectively, from
their Figure 5.  The Castellano et al.\ (2016b) catalogs become
significantly incomplete ($>$50\%) at $>$28.5 mag even adopting point
source profiles.  Livermore et al.\ (2017) do not provide a figure
showing their estimates of completeness at $\sim$29 mag, but making
similar assumptions about the intrinsic sizes of $z\sim6$ galaxies to
what they report in their paper, the completeness would not appear to
be higher than 15\% at $>$29 mag.

In general, we would expect errors in the selection volumes to become
large faintward of 28.5 mag and especially at $>$29 mag.  We
illustrate this by plotting the selection volumes we estimate using
different assumed intrinsic half-light radii in
Figure~\ref{fig:liver1} for a $z\sim6$ selection.  The completeness
corrections that need to be applied in the derivation of the UV LF
from the HFF data often exceed a factor $\sim$5$\times$ at $>$29 mag.
This correction is highly sensitive to assumptions about the galaxy
sizes.  Clearly, it can be very risky to include such sources in
estimates of the LF (especially when the size distribution of sources
is not yet clear).\footnote{For the most accurate results, of course,
  the selection volumes and completeness must be derived from a
  forward modeling procedure that incorporates the impact of a model
  faint-end slope and photometric scatter.}

As the Atek et al.\ (2015) and Castellano et al.\ (2016b) studies
include only modest numbers of sources fainter than 28.5 mag, we would
expect their LF results to be less impacted by the size distribution
they assume.  For Alavi et al.\ (2016), no presentation of the
apparent magnitude distribution is provided, so its importance is not
clear, but it is repeatedly emphasized as a large uncertainty in their
derived LF results.

One case where the size distribution is likely to have a large impact
is for the Livermore et al.\ (2017) study.  Their samples extend to
$\sim$29.3 mag.  $\sim$30\% of their sample lies faintward of
$\sim$29 mag where the completeness corrections are large and
uncertain (see histogram in the lower panel of Figure~\ref{fig:liver1}
and compare with the completeness corrections shown in the top panel).

Livermore et al.\ (2017) assume that faint sources have a median
half-light radius of $\sim$90 mas (0.5 kpc at
$z\sim6$).\footnote{Interestingly, the faintest source in the
  Livermore et al.\ (2017) sample points to a source size smaller than
  they assume is typical for the faint population.  Fitting this
  galaxy with the profile-fitting software \textsc{galfit}, we find a
  half-light radius of $<$70 mas in the image plane.  Taking a
  magnification factor of $\sim$50 from the Ishigaki et al.\ (2016)
  \textsc{glafic} (Oguri 2010) model, this translates to the source
  having an intrinsic half-light radius of $<$10 mas, consistent with
  what we are finding in this paper (and small compared to 90 mas).}
With the smaller sizes for very faint galaxies implied by the current
results ($<$20-30 mas) we would expect the volume density of faint
sources that Livermore et al.\ (2017) derive to be significantly
higher than what we derive.  We would also expect their faint-end
slope results to be biased towards steeper values (as in
Figure~\ref{fig:lf6size}.)

As we consider such an assessment of the Livermore et al.\ (2017) LF
results, it is useful to understand a little about how these results
were derived.  The faint-end slope $\alpha$ that would be obtained
from a fit to the individual points in their $z\sim6$ LF points is
$\sim-2.3$ (see their Figure 10).  Livermore et al.\ (2017) then apply
an Eddington bias correction that reduces this slope.  The Eddington
bias correction Livermore et al.\ (2017) apply is quite large in
specific magnitude ranges, i.e., $\sim$0.5-0.8 mag, and results in a
faint-end slope of $\sim-2.1$.  [This estimate for the applied
  Eddington bias correction is derived based on the horizontal offset
  between the solid and dotted purple lines in their Figure 11.]
Based on our own simulations we cannot justify such large Eddington
bias corrections and suggest that $\sim$0.075 mag may be more
appropriate.  This would imply more minimal corrections to the LF.
Unless we have missed something, it would appear that Livermore et
al.\ (2017) applied a correction which was too strong in reducing
their faint-end slope results from $\sim-$2.3 to $\sim-$2.1.

With this background, we now present in Figure~\ref{fig:liver2} a
comparison of the Livermore et al.\ (2017) $z\sim6$ LF with other
determinations of this LF from Atek et al.\ (2015), Bouwens et
al.\ (2015), and Bouwens et al.\ (2017).  No correction is made for
Eddington bias to ensure all four LF results are treated similarly.

What is striking about this comparison is that it demonstrates again
(as in Figure~\ref{fig:lf6size}) the crucially important role played
by using the actual sizes of very faint galaxies when deriving LF
constraints (and the need for special caution around the completeness
limit because of the large corrections required).

\subsection{Implications of these results for state-of-the-art selection 
volume methods}

Results from this paper suggest a steep size-luminosity relation at
high redshift, with faint galaxies being very small in physical size,
having morphologies that are much more akin to point sources than we
previously expected (5-10 mas intrinsic half-light radii, equivalent
to 20-50 pc at z$\sim$6 and 40-80 pc at z$\sim$2-3).

For such small source sizes, we would expect that the most discernible
impact of lensing to be on the total fluxes.  The impact of shear is
less important.  This is of enormous convenience to know if one is
running simulations, since it means that we can perhaps model galaxies
as if they were point sources, and for many applications it appears
this produces reasonable results.  The advantage is that simulations
can be run much more similarly to those for blank fields such as the
HUDF.\footnote{Given that real galaxies must have non-zero sizes, it
  is clear that this is an approximation and therefore it must break
  down in certain regimes.  However, the point we make in the current
  paper is that one can obtain surprisingly reasonable selection
  volume estimates by simply treating galaxies as point sources.}

As the current observational results have illustrated, there appears
to be no strong empirical motivation to accurately model the impact of
lensing on the spatial profiles or morphologies of galaxies for
estimating completeness.  Even after considering every faint $z=2$-8
source in the high magnification regions behind 4 HFF clusters, we
recover essentially equal surface densities of galaxies in regions of
both low and high shear, as illustrated by Figure~\ref{fig:compshear}.
No statistically significant trend is present vs. shear factor.  Along
similar lines, a stack of faint sources predicted to be highly sheared
reveals little elongation along the axis of maximum shear.  This
suggests that the faintest galaxies are not only very compact, but
also show no evidence for diffuse, lower-surface brightness structure.

If we cannot (after considerable effort) detect such spatial
distortion effects from lensing in our total sample of faint $z=2$-8
galaxies identified behind the first 4 HFF clusters, it suggests that
such effects may only have modest impact on the visibility of sources
behind lensing clusters, and one could potentially ignore the impact
of lensing in running selection volume simulations to estimate the
luminosity function.

We emphasize that the recommendation we provide above is based on the
study of very faint sources with absolute magnitudes $>-16.5$ mag.
For larger, brighter sources, lensing shear is almost certainly more
important, and detailed simulations (which include lensing
transformations) are likely required to derive accurate selection
volume and completeness estimates, as recommended earlier by Oesch et
al.\ (2015).

\section{Summary}

In this paper, we present an entirely new approach for deriving
constraints on the size distribution of extremely faint galaxies seen
behind the HFF clusters.  Strong constraints on this size distribution
are essential for estimating the efficiency (or completeness) with
which we can find such faint galaxies behind the HFF clusters and
hence obtaining accurate constraints on their prevalence.

The approach keys on the idea that highly-magnified galaxies should be
significantly easier to find in regions with low shear than high
shear.  Large intrinsic sizes would result in the largest differences
between the two shear regions, while small, almost point-like sizes
would result in essentially no differences in the observed surface
densities in the two regions.

Using sophisticated image construction and source recovery
simulations, we quantify how the selection efficiency of galaxies
would depend on the predicted shear for S/N appropriate for the HFF
clusters, for a variety of different assumptions about the intrinsic
sizes.

Taking advantage of a large sample of 87 high-magnification
($\mu=10$-100) $z\sim2$-8 galaxies identified behind the first four
HFF clusters, we look at how the surface density of high-magnification
$z\sim2$-8 galaxies depends on the predicted shear.  Remarkably, we
find that our observed samples show no statistically-significant
dependence on shear.  Leveraging our simulation results to interpret
this observational finding, we conclude that extremely faint
($\sim-$15 mag) galaxies have intrinsic half-light radii less than 30
mas and 60 mas (87\% and 99\% confidence, respectively).

The constraints we can set on the overall size distribution weakens if
we consider galaxies to have a range of sizes, due to our tendency to
preferentially select the smallest galaxies.  For a lognormal size
distribution with 0.3 dex scatter, we infer that the median intrinsic
half-light radius is no larger than 30 mas and 60 mas (74\% and 99\%
confidence, respectively).  

As a basic check on the size constraints we obtained using this new
shear-based technique, we also examined the spatial profiles of 26
intrinsically-faint ($>-16$ mag) $z=5$-8 galaxies expected to be
stretched by $>$10$\times$ along one dominant shear axis.  We compared
the spatial profile of these galaxies with what one would expect for
various assumptions about the intrinsic half-light radius and also
using the CATS lensing model.

Amazingly, sources showed essentially no evidence for spatial
extension along the major shear axis.  This was true both on an
individual basis -- with suggestive half-light radii of $\lesssim$3-10
mas -- and after stacking the spatial profile of many sources expected
to be elongated by $>$10$\times$ along a dominant shear axis.  Our
stack results suggest intrinsic half-light radii of $\sim$4 mas for
the faint galaxies, which corresponds to $\sim$20 pc at $z\sim6$.
Remarkably, this is even smaller than many $z\sim0$ GMCs or
star-forming clumps seen in galaxies from $z=0$ to 4 (Kennicutt et
al.\ 2003; Livermore et al.\ 2012, 2015).

These results are of enormous importance for determinations of the
faint end of the $UV$ LF at $z\sim2$-9, as they allow for a proper
quantification of the probable selection efficiencies and volumes for
faint $z\sim2$-9 galaxies.  Without such a quantification, the
faint-end slope $\alpha$ determinations need to factor in large
uncertainties resulting from the lack of knowledge of the size
distributions for faint galaxies.  In such cases, assumptions are made
about the sizes that can lead to incorrect determinations of the
volume density and LF shape.

In this context, it is clear that many recent LF results at $>$$-17$
mag are significantly impacted by the typically large sizes of
$\sim$100 mas assumed for faint sources.  For example, accounting for
the much smaller sizes found here, we would expect dramatically lower
volume densities for very faint sources and also flatter faint-end
slopes than has been reported in many recent works.  The differences
can be as much as $\Delta\alpha\sim0.2$-0.3 or larger, which makes a
very large difference to the UV luminosity density computed from
galaxies in the reionization epoch (Figure~\ref{fig:lf6size}).  In
Bouwens et al.\ (2017), we present determinations of the faint-end
slope $\alpha$ of the $z\sim6$ LF using the present size constraints
and find $\alpha=-$1.92$\pm$0.04.

The steep size-luminosity relations suggested by the present results
have important implications for the simulations that must be run to
estimate the selection volumes behind lensing clusters.  Indeed, if we
take as representative our current results where almost all faint
sources are small, i.e., $<$20 mas, we can arrive at surprisingly
accurate estimates of the selection volume, simply by treating faint
($>$30 mag) galaxies as point sources.  This simplification is of
great value to anyone deriving the needed selection volumes, since it
means that one can, for all practical purposes, ignore the impact of
lensing on the spatial profiles or morphologies of galaxies and simply
make use of selection volume simulations such as we use on the HUDF.

In drawing the present conclusions on source sizes, we remind our
audience that we exercise a significant reliance on the lensing models
that have been made publicly available over the HFF clusters.  If
there exist any large systematic inaccuracies in those models, it
would impact the conclusions drawn in this paper.

An important priority for future work would be to extend the current
study to brighter sub-$L^*$ sources where the impact of lensing can
clearly be seen in the morphologies and then to study the transition
from extremely faint galaxies behind the HFF clusters to galaxies
which are slightly brighter.

While it clearly remains desirable to enhance the sample size and to
derive more accurate constraints on the size distribution as a
function of magnitude for high redshift galaxies at $z\gtrsim2$, the
bottom line from the present study indicates quite strongly that for
high redshift galaxies fainter than 30 mag, the half-light sizes
should be taken to be $<$30 mas ($\sim$200 pc) with the likelihood
that half-light sizes around 5-10 mas (30-60 pc) may well be quite
common.\\

\acknowledgements

We acknowledge the support of NASA grant HST-AR-13252, NASA grant
HST-GO-13872, NASA grant HST-GO-13792, NWO vrij competitie grant
600.065.140.11N211, and NWO TOP grant TOP1.16.057.

\appendix

\section{A.  Size Measurements for Brighter sub-L$^*$ Galaxies and Comparisons with Sizes Measurements of Field Galaxies}

In \S5, we attempted to obtain direct constraints on the sizes of
sources in the high magnification regions.  Amazingly, the half-light
radii we inferred for faint sources from direct size measurements was
$\sim$4 mas for $z\sim5$-8.  4 mas is equivalent to physical sizes of
25 pc for individual galaxies for very faint galaxies.

Given the possibility that such extreme size inferences are the result
of systematic errors in the lensing models, it is useful to attempt to
constrain the sizes of more luminous sources where the size
distribution is more well established from studies over fields like
the HUDF.

\begin{figure*}
\epsscale{0.9}
% /Users/bouwens/magmodel/xdf
\plotone{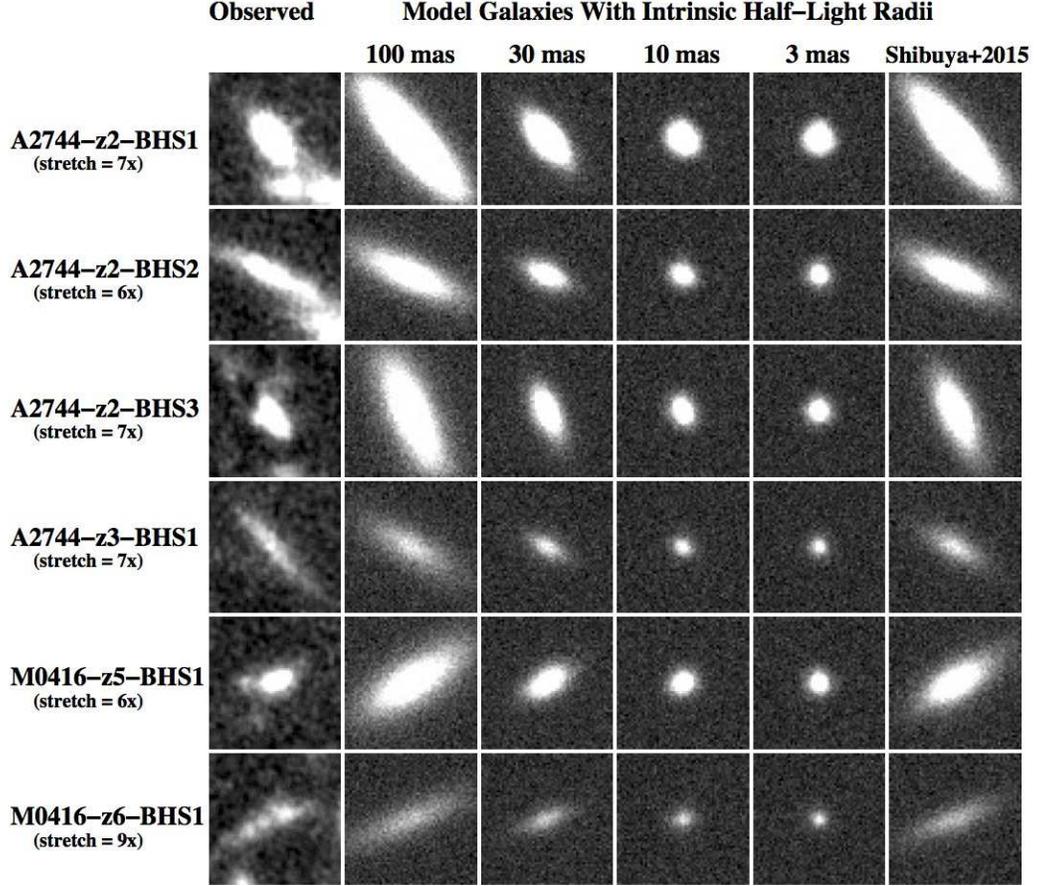}
\caption{Similar to Figure~\ref{fig:indprof} but for $z\sim2$-8
  galaxies with higher intrinsic luminosities.  The intrinsic apparent
  magnitudes for the plotted sources range from $\sim$27 mag to
  $\sim$29.4 mag, equivalent to $M_{UV,AB}\sim-20 mag$ to
  $M_{UV,AB}\sim-17.6$ mag, according to the median magnification
  model.  The observed spatial profiles of the sources are in
  reasonable agreement with that expected based on the approximate
  size-luminosity relations given in Shibuya et al.\ (2015), with
  intrinsic half-light radii ranging from 30 mas to 100
  mas.\label{fig:indprofzb}}
\end{figure*}

We therefore repeat the exercise from \S5 but this time only
considering those sources with intrinsic magnitudes in the range 27.0
mag to 29.4 mag, where size constraints are already available from
fields like the HUDF.  As in \S5, we again treat sources as having an
approximate circular profile, with half-light radii of 3 mas, 10 mas,
30 mas, and 100 mas, as well as source sizes given by the Shibuya et
al.\ (2015) size-luminosity relations.  The results are presented in
Figure~\ref{fig:indprofzb} for the six well-separated sources in the
magnitude range in question.

As should be clear from Figure~\ref{fig:indprofzb}, the observed
sources generally have sizes and spatial profiles that are reasonably
consistent with expectations for $\sim$28 mag galaxies.  We can look
at this comparison more quantitatively by using \textsc{galfit} to fit
the profiles of the sources shown in Figure~\ref{fig:indprofzb}.
Fitting to each of the sources plotted here accounting for the impact
of lensing as in \S5.1, we measure a mean intrinsic half-light radius
of 0.35 kpc vs. an expected mean half-light radius of 0.42 kpc, which
is excellent agreement overall.

\end{document}